\documentclass{article}

\usepackage{arxiv}

\usepackage[T1]{fontenc}    
\usepackage{hyperref}       
\usepackage{url}            
\usepackage{booktabs}       
\usepackage{amsfonts}       
\usepackage{nicefrac}       
\usepackage{microtype}      
\usepackage{lipsum}		
\usepackage{graphicx}
\usepackage{natbib}
\usepackage{doi}

\usepackage{makecell}
\usepackage[utf8x]{inputenc}
\usepackage{tfrupee}
\usepackage{adjustbox}
\usepackage{longtable}
\usepackage{tcolorbox}
\tcbuselibrary{skins,breakable}
\usepackage{enumitem}
\usepackage{hyperref}   

\title{Unpacking ``Personal'' Health Informatics for \emph{Proactive Collective Care}}
\renewcommand{\shorttitle}{Designing for \emph{Proactive Collective Care}}

\author{ \href{https://orcid.org/0000-0001-5787-4858}{\includegraphics[scale=0.06]{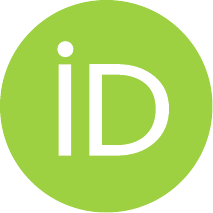}\hspace{1mm}Shyama Sastha~Krishnamoorthy Srinivasan}\\
	IIIT-Delhi\\
	New Delhi, India\\
	\texttt{shyamas@iiitd.ac.in} \\
    \And
    \href{https://orcid.org/0000-0002-0286-6997}{\includegraphics[scale=0.06]{orcid.pdf}\hspace{1mm}Mohan~Kumar}\thanks{Both authors are advisors of this work.}\\
	RIT, NY\\
	Rochester, New York, US\\
	\texttt{mjkvcs@rit.edu} \\
	\And
	\href{https://orcid.org/0000-0003-2152-1027}{\includegraphics[scale=0.06]{orcid.pdf}\hspace{1mm}Pushpendra~Singh}\footnotemark[1]\\
	IIIT-Delhi\\
	New Delhi, India\\
	\texttt{psingh@iiitd.ac.in} \\
}

\renewcommand{\shorttitle}{Designing for \emph{Proactive Collective Care}}

\hypersetup{
pdftitle={Unpacking ``Personal'' Health Informatics for Proactive Collective Care},
pdfsubject={cs.HC, cs.CY},
pdfauthor={Shyama Sastha~Krishnamoorthy Srinivasan, Mohan~Kumar, Pushpendra~Singh},
pdfkeywords={Ecology of tracking, Personal Health Management, Collective Care},
}

\begin{document}
\maketitle

\begin{abstract}
	Care is primarily a collective phenomenon, with a practice that involves sharing health and wellbeing information within a trusted ``care circle'' of family members and companions for sensemaking, interpretation, decision-making, and follow-through. However, current digital health tools and information systems are designed for individuals and primarily intended for Personal Health Informatics (PHI). This mismatch between collective practice and individualistic design creates new challenges for the proactive use of such systems in care settings and limits adoption, sustained engagement, and meaningful use. To examine how people practice collective care and how (if) they perceive, adopt, and integrate PHI systems for proactive care, we conducted a sequential mixed-methods study. Through an initial survey (n=87) and semi-structured interviews (n=22), we found that their practices involve collectively understanding, analyzing, and sensemaking health information. However, we also found that their use of existing systems to support such practices is constrained by factors at personal, relational, technological, and structural levels that evolve over time. To address these constraints and explore redesigning PHI toward ``Collective Health Informatics'', we conducted eight stakeholder interview sessions with 12 participants, a follow-up survey (n=114), and six co-design workshops with 12 participants to understand the dynamics required in collective settings while retaining agency. Using a design probe evaluation (n=38), we refine a design vision for coordinated, trustworthy action across such care relationships. Our findings motivate \textit{CC-Proact}, an operational map that translates ecological influences into three design levers: \textit{Agency}, \textit{Elicitation}, and \textit{Engagement}. Using this map, our work empirically examines collective care practices and offers six design recommendations to build responsible systems that proactively support collective care.
\end{abstract}


\section{Introduction}
Care is rarely a solitary act; it is a shared social practice woven into the quiet, often invisible labor of daily life\footnote{In CSCW literature, care is understood through Star and Strauss's~\cite{starLayersSilenceArenas1999} concept of ``articulation work.'' It is the uncalculated, background social practices and invisible labor necessary to sustain cooperative human systems and daily routines.} Whether it is a spouse noticing a partner’s unusual fatigue after a workday or a daughter remotely checking her otherwise healthy father’s heart rate, care thrives in the collective. It needs a ``care circle''\textemdash a trusted network of family, friends, companions, and healthcare professionals who collectively participate in sensemaking of health signals, negotiate treatment, and offer emotional scaffolding. While self-care is a vital subset of this spectrum, broader care is fundamentally relational. It involves more than one body and more than one mind. Prior work in the HCI/CSCW community has focused on individual aspects of such relational care, including sociocultural dimensions~\cite{bhatsocioculturaldimensionstracking2020}, shared responsibility~\cite{chasharedresponsibilitycollaborative2024}, and decision-making~\cite{blandfordhcihealthwellbeing2019}. To explicitly focus on the overall cooperative aspect of such care, we define \textbf{collective care} as \textit{a culturally grounded, trust-based system of shared responsibility for health and wellbeing that shapes information flow, decision-making, and support within a care circle.} 

While many current-day digital health tools, such as smartwatches and wearables, primarily promise to support proactive care\footnote{While preventive care centers on formal clinical risk detection, proactive care emphasizes daily, continuous social sensemaking and early collaborative interventions within a care circle before clinical thresholds are breached.}~\cite{waldmanandterzicproactive} by helping people monitor, interpret, and act on their health signals from their daily lives, they typically have an individualistic design focusing on Personal Health Informatics (PHI). They typically follow the stage-based~\cite{10.1145/1753326.1753409} or lived informatics~\cite{epsteinlivedinformaticsmodel2015} models, which remain useful but offer limited guidance on collective practices and often rely on ad hoc workarounds. For instance, prior examples on health systems for continuous monitoring~\cite{stuartwearabledevicescontinuous2022, 10.1145/3749503} were designed around an individual user who is expected to understand, interpret, and act on the data independently. This design assumes an island of agency that simply does not exist for many. This assumption breaks down when health management is distributed among a care circle, especially when users rely on others for interpretation, prompting, or follow-through.

Prior work in health informatics for collaborative settings further emphasizes that health data use is socially situated and that design must account for more than a single autonomous user. For instance, prior research on family informatics highlights the affective and behavioral dynamics of family interactions~\cite{leefamilyscopevisualizingaffective2024}, while family-centered and inter-generational tracking studies demonstrate how health data is interpreted and acted upon across members of a household~\cite{10.1145/2998181.2998362, bindaintergenerationalsharinghealth2017}. Work on care networks further emphasizes the role of social support in coordinating care for older adults~\cite{1316814}, and studies of caregiving practices show how health data is collaboratively produced, interpreted, and used in contexts such as chronic illness and mental health management~\cite{10.1145/2998181.2998303, 10.1145/3025453.3025843}. Ecological approaches extend this view by situating the use of health systems within broader networks of relationships and infrastructures that shape care practices~\cite{murnanepersonalinformaticsinterpersonal2018}. However, these studies remain fragmented across specific contexts, relationships, and conditions, offering limited guidance on how such systems should support shared interpretation, shared responsibility, and coordinated action in everyday collective care settings. 

The gap causing this breakdown becomes especially visible in a context like urban India, where universal tensions between individualistic design and collective reality are evident. Prior work has already shown that the acceptance, adoption, and sustained use of health technology is shaped not only by device availability but also by a myriad of factors such as fragmented technology ecosystems~\cite{karusalaunsettlingcareinfrastructures2023}, uneven digital health literacy~\cite{okoloifiteasy2024}, trust in sensors and platforms~\cite{paideterminantsmobilehealth2020}, everyday role of care circles in health decisions~\cite{10.1145/3025453.3025843, bhatsocioculturaldimensionstracking2020}, and other inconveniences~\cite{patilfactorsaffectingusage2022}. Rather than treating these conditions as background noise, we view them as part of the sociotechnical environment that determines whether redesigning such systems can, in practice, proactively support collective care.

To understand how to support such settings, we examine how care is currently practiced and PHI is (or is not) used in those practices, identify gaps that require workarounds, and ideate the design changes needed to support collective care. For this work, we focused on urban India as an example of such contexts, where collective care practices are common yet remain underexplored in health system design, amid the increasing adoption of proactive care since COVID-19 \footnote{\href{https://www.pib.gov.in/PressReleaseIframePage.aspx?PRID=2094604&reg=3&lang=2}{Press Information Bureau (PIB) release from India's Ministry of Health and Family Welfare, with letter from World Economic Forum (WEF)}}.

Accordingly, we ask the following research questions: 
\begin{itemize}
    \item[\textbf{RQ1}] How do people in urban settings currently engage in collective care?
    \item[\textbf{RQ2}] How might current systems be redesigned to proactively support collective care practices?
\end{itemize}

To answer \textbf{RQ1}, we sought to understand people's current care practices, including their use of PHI for those practices. To that end, we surveyed 87 participants and conducted 22 semi-structured interviews to examine awareness, use, motivations, and barriers related to PHI, as well as their current practices in everyday settings. This revealed a persistent gap between current system designs and collective care practices around sharing, family mediation, and responsibility, as well as patterns of PHI use, reconfirming barriers/opportunities. To further investigate how existing systems might better support collective care practices, we conducted 8 stakeholder interviews and 6 co-design workshops to explore how actors participate, what they expect, and what they are willing to \textit{share}. We then evaluated a Figma\footnote{https://www.figma.com/} prototype with 38 participants to explore how such health systems might better support coordinated, trustworthy action, addressing \textbf{RQ2}.

\begin{figure*}[ht]
    \centering
     \includegraphics[width=.95\linewidth]{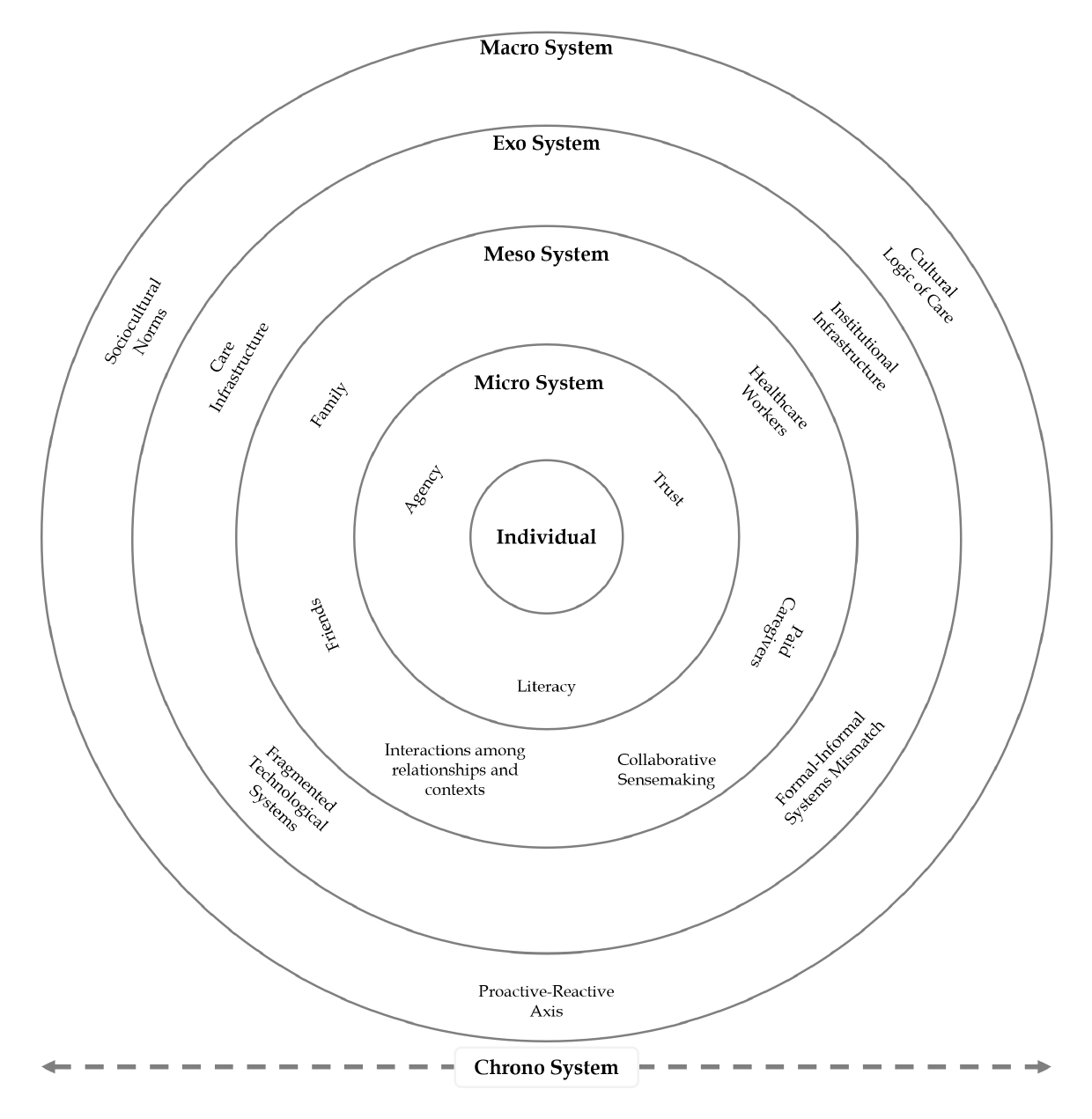}
     \caption{\label{Figure 0} Ecosystems of Collective Care based on Bronfenbrenner's Ecological Systems Theory~\cite{bronfenbrennerdevelopmentalresearchpublic1974}.}
\end{figure*}

Urie Bronfenbrenner’s ecological systems theory (EST)~\cite{bronfenbrennerdevelopmentalresearchpublic1974} describes how human experiences are shaped by multiple interconnected ecosystems, including interpersonal, organizational, and broader contextual environments. We use this theory to frame proactive collective care as a socio-technical process shaped by interactions across these ecosystems (Figure~\ref{Figure 0}). Across the study, participants sought to share health informatics within the care circle, obtain help interpreting them, and coordinate actions based on them. However, these practices were often unsupported by existing systems, requiring ad hoc workarounds such as manual sharing, external reinterpretation tools, and reliance on trusted intermediaries. Adoption and use were further constrained by ecosystem-level factors, including low data confidence, limited interface accessibility, fragmented tools, and the need to contextualize health information through relationships within the care circle. These findings indicate that supporting proactive collective care requires enabling coordinated, trustworthy action across the entire care circle (including the care subject) while accounting for all ecological influences, rather than merely improving individual tracking or sensemaking.

These findings motivated the development of CC-Proact, our primary contribution of this work. \textit{The CC-Proact operational map is} a conceptual design map whose design is grounded in the empirical accounts of everyday practices by participants, operationalizing the shift to proactive collective care by translating complex ecological influences into three actionable design levers: Agency, Elicitation, and Engagement. Using this design map, we translate these levers into six design recommendations, providing actionable guidance for building responsible systems that honor the interdependence of relationships to support proactive collective care.
\section{Related Work}

Our research is motivated by prior work on health informatics, care networks, and the social organization of health data. This literature points to a gap in design guidance for proactive collective care, where shared interpretation, selective disclosure, and coordinated action must be supported simultaneously. To unravel this, we first consider how an individualistic conceptualization is detrimental in collective settings, then turn to work that reframes health informatics as relational and collectively managed, and finally to scholarship that shows the different ecosystems that shape this context and the tensions they create.

\subsection{The ``Quantified Self'' and Its Boundaries}
 PHI has been extensively studied as a means of supporting individuals in understanding and managing their health through data. Foundational models, such as the stage-based model of personal informatics~\cite{10.1145/1753326.1753409}, characterize tracking as a process spanning 5 stages (preparation $\rightarrow$ collection $\rightarrow$ integration $\rightarrow$ reflection $\rightarrow$ action) and identify barriers that arise at each stage. Subsequent work has extended this view through lived informatics~\cite{rooksbypersonaltrackinglived2014, epsteinlivedinformaticsmodel2015}, emphasizing that tracking is not a linear pipeline but an ongoing, situated practice shaped by evolving goals, interruptions, and changing contexts. Together, these models have been instrumental in shifting PHI from simple data logging to a richer understanding of how people plan, track, reflect on, and act on personal health data.

For instance, a substantial body of work has examined motivations for self-tracking and challenges that arise in practice. People track for a range of reasons, including behavior change~\cite{intille2004ubiquitous}, curiosity~\cite{rapppersonalinformaticseveryday2016}, maintaining records~\cite{zaheremergingrolewearable2024}, workplace wellness~\cite{10.1145/3025453.3025510}, and managing specific health conditions~\cite{10.1145/3314409}, and these motivations shape how such tracking systems are accepted, adopted, sustained, or even abandoned~\cite{shinwearableactivitytrackers2019, samhaleimpacttrustinternet2022, lazarwhyweuse2015}. Across the stages from the stage-based model, barriers emerge from mismatches between tools and user goals~\cite{10.1145/2750858.2805832}, difficulties in integrating heterogeneous data~\cite{10.2196/31618}, limited perceived utility~\cite{kruzanperceivedutilitysmartphone2023}, and challenges in interpreting data~\cite{lustudyrelationshiptechnology2024, 10.1145/3314409} or translating insights into action~\cite{10.1145/3749503, nunesselfcaretechnologieshci2015}. Prior systems have sought to address these challenges by supporting end-to-end tracking processes either through focusing on specialized contexts (such as fall detection)~\cite{basticosimultaneousrealtimehuman2022}, or enabling customization of tracking variables for specific goals~\cite{brunexplorationelectrochromicscalm2022}, or scaffolding individualistic goal-directed tracking~\cite{zhaogrowmeexploring2024, 10.1145/3749503}. However, this research largely assumes that the individual tracker is solely responsible for defining goals, interpreting data, and acting on insights, even when domain knowledge or contextual understanding may be limited.

More recent work has begun to acknowledge that health informatics is not entirely individual. Studies highlight how tracking is embedded in everyday routines, influenced by social context, and sometimes supported by external actors such as family~\cite{10.1145/2998181.2998303, chasharedresponsibilitycollaborative2024}, clinicians~\cite{10.1145/3411764.3445587, 10.1145/3544548.3581251}, or peers~\cite{haquebuttcallme2024, 10.1145/3710978}. However, these contributions typically extend individual models rather than modeling the collective as a whole, accounting for all factors involved. Other people involved are often treated as sources of feedback, expertise, or responsibility rather than as collaborators in ongoing care. As a result, sharing is commonly treated as an output of personal tracking, through raw data, generic summaries, information dashboards, or reports\textemdash rather than as a core mechanism for interpreting and acting on health data.

Taken together, this literature provides a strong foundation for understanding how individuals engage with health data across stages of tracking, and how systems can better support planning, reflection, and action. At the same time, it leaves open a critical question for PHI design: How should systems support health management when interpretation and action are not centered on a single user, but distributed across a set of people who collectively make sense of and respond to health information?

\subsection{Collaborative Sensemaking \& Shared Agency}

A growing body of work in HCI\textemdash spanning care networks and family informatics\textemdash reframes health data not as a purely personal resource, but as something produced, interpreted, and acted upon within relationships ~\cite{1316814, leefamilyscopevisualizingaffective2024, 10.1145/3025453.3025843, ugargolfamilycaregivingolder2018, gustafssoninformalcaregivingperspectives2022, creaserexploringfamiliesacceptance2022, 10.1145/3534614, 10.1145/2675133.2675200, karnatakitevengiving2023, bindaintergenerationalsharinghealth2017, 10.1145/3240925.3240936, 10.1145/3421937.3422018, 10.1145/3706598.3713596}. Early work on care networks~\cite{1316814} positioned individuals as embedded within systems of support, including family members, caregivers, and clinicians, highlighting that health management often involves coordination among multiple actors rather than isolated decision-making. Building on this, family informatics research~\cite{leefamilyscopevisualizingaffective2024, 10.1145/3025453.3025843} has shown that health tracking often becomes a shared activity, in which family members monitor one another's health and wellbeing, exchange observations, and use data to support collective goals such as supporting a subject of care~\cite{ugargolfamilycaregivingolder2018, gustafssoninformalcaregivingperspectives2022, creaserexploringfamiliesacceptance2022}, managing chronic conditions~\cite{10.1145/3534614, 10.1145/2675133.2675200, karnatakitevengiving2023}, or maintaining routines~\cite{bindaintergenerationalsharinghealth2017, 10.1145/3240925.3240936, 10.1145/3421937.3422018, 10.1145/3706598.3713596}.

Subsequent studies deepen this relational view by examining how health data is used in practice within families and care settings. Work on intergenerational~\cite{bindaintergenerationalsharinghealth2017, 10.1145/3240925.3240936, 10.1145/3421937.3422018} and family-based tracking~\cite{10.1145/2998181.2998362} demonstrates that shared health data support not only awareness but also communication, accountability, and negotiation of responsibilities. In caregiving contexts, as observed by Kaziunas et al.~\cite{10.1145/2998181.2998303}, tracking and record-keeping are often distributed tasks, with different actors contributing observations, maintaining logs, and interpreting trends over time. Research on manual tracking~\cite{10.1145/3025453.3025843} in caregiving for those with mental health support needs further shows that these practices are not only informational but also social, shaping how care is coordinated and how relationships are maintained. Across these studies, health data is less an individual artifact and more a medium through which care is organized.

This perspective is further reinforced by observations on health datafication~\cite{bhatwearehalfdoctors2023, thakkarwhenmachinelearning2022}, which highlight the social and emotional labor involved in producing and using health data. They show the importance of ``boundary actors'' who connect patients with clinicians, manage information across settings, and coordinate care across formal and informal systems. To support such behavior, data must be made legible to others, contextualized within lived experience, and translated into decisions that are acceptable within a given relationship. Trust, reciprocity, and role differentiation become central: who is allowed to see what, who is responsible for acting, and how disagreements or uncertainties are resolved. These concerns extend beyond simple sharing mechanisms, pointing instead to the need for systems that can support ongoing negotiation and coordination within care relationships.

While these prior works convincingly establish that health data is relational, it primarily treats information sharing primarily as an output or summary report rather than an ongoing mechanism for collaborative sensemaking; it remains fragmented across dimensions, such as specific health conditions, relationships, and caregiving contexts; and provides less consolidated guidance for designing systems that proactively support collective care in everyday settings. In particular, there is limited articulation of how PHI systems might simultaneously support shared interpretation, selective disclosure, and coordinated action across a care circle. This gap motivates a shift from viewing sharing as an output of personal tracking toward understanding it as a central mechanism through which health data becomes meaningful and actionable.

\subsection{Mapping Health Information Ecologies}

Prior work in HCI has shown that sociotechnical conditions deeply shape the acceptance, adoption, and sustained use of digital health practices. Hoque et al.~\cite{hoqueculturalinfluenceadoption2015} and Zhang et al.~\cite{zhangimpactmoderatingeffect2022} show the impact of cultural and national influences on digital health practices by cross-examining technology acceptance through Hofstede's cultural dimensions. Specifically, the former demonstrates how rigid societal hierarchies can prioritize functional utility over user experience to drive adoption in developing regions, while the latter positions national culture as a macro-level gatekeeper, in which individualistic autonomy catalyzes technology acceptance and risk-averse uncertainty avoidance paralyzes it. Singh et al.~\cite{singhpresentfuturehealth2022} present observations on data collection and processing to support critical health infrastructure, showing how time and health economics influence these practices, extending beyond culture. Seelam et al.~\cite{seelamfactchecksaretop2024} examine the importance of awareness and technology's reach for the utility of critical health practices such as fact-checking, further highlighting the impact of media and information reach. Building on these, Karusla et al.~\cite{karusalaunsettlingcareinfrastructures2023, karusalaspeculatingcareworkercentered2023} shed light on the impact of overburdened health workers and the counterintuitiveness of scaling up care support, noting its effects on care infrastructure. Prior works on literacy~\cite{auldhealthliteracyhealth2020, krepsrelevancehealthliteracy2017, mackerthealthliteracyhealth2016, meyersparentsusetechnologies2020}, and trust~\cite{10.1145/3530190.3534824, zhanhealthcarevoiceai2024, okoloifiteasy2024} further shed light on how an individual's behavior in digital health use is affected by internal (self) and external (family, friends, government policies, technology availability) influences. These observations reinforce the role of family and trusted others as mediators of health information, where decisions are often made collectively rather than independently, and show that effective care practices require more than just the actors involved. 

Taken together, these observations consistently highlight that care is rarely an individual activity, but is instead influenced/mediated by family, friends, society, infrastructure, culture, government, and time for acceptance, adoption, interpretation, decision-making, and follow-through. This positions health informatics not as something owned and acted upon by a single individual, but as something that moves through and is maintained within an interlinked set of care ecosystems. Prior works examining PHI within a network of health information ecologies, integrating formal health systems and informal social contexts for mental health~\cite{murnanepersonalinformaticsinterpersonal2018, tachtlerunaccompaniedmigrantyouth2021, waniunresttraumastays2024}, chronic conditions~\cite{bhatsocioculturaldimensionstracking2020}, fertility~\cite{costafigueiredohealthdatafertility2021}, and many more mirror the different ecosystems described in Bronfenbrenner's EST~\cite{bronfenbrennerdevelopmentalresearchpublic1974}. Similarly, we can map these distinct influencers of care into Micro, Meso, Exo, and Macro systems, with the Chrono system showing how they evolve over time, to identify opportunities and barriers within each system and proactively support collective care practices.

Systems that presume a single, autonomous user with stable access, sufficient literacy, and independent decision-making authority may not align with how care is actually practiced. This work provides a lens for examining how redesigning systems might better support collective care in environments where health management is collective, mediated, and constrained.
\begin{figure*}[ht]
    \centering
     \includegraphics[width=.95\linewidth]{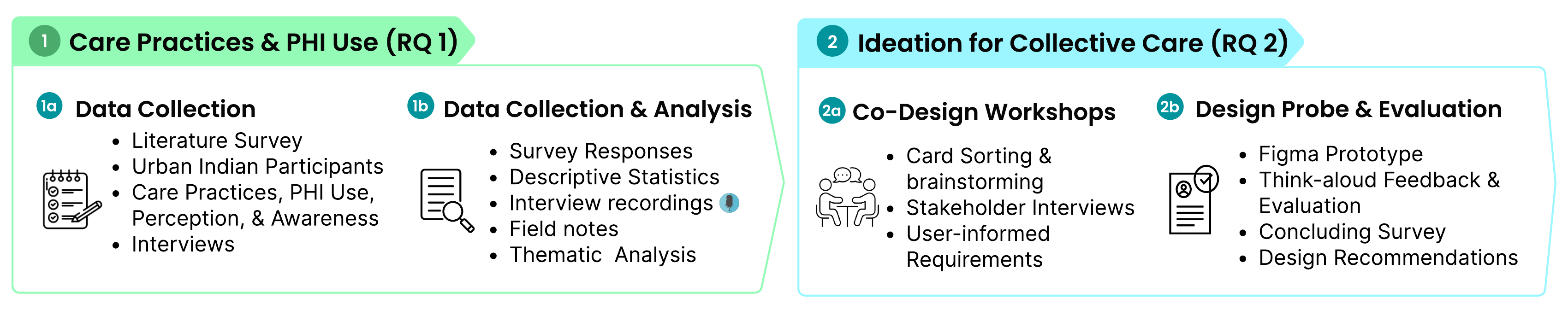}
     \caption{\label{Figure 1} Workflow of the sequential Study. The sequence follows Surveys $\rightarrow$ Interviews $\rightarrow$ Analysis/Findings for RQ2 $\rightarrow$ Co-design Workshops + Stakeholder Interviews $\rightarrow$ Prototype Design \& Evaluation + Sharing Feature Survey $\rightarrow$ CC-Proact Design Map + Recommendations.}
\end{figure*}

\section{Methods}

We adopted a sequential mixed-methods design, in which qualitative and quantitative components are chronologically connected so that outcomes from one phase explicitly inform the purpose, sampling, or procedures of the next as explained by Ivankova et al. ~\cite{ivankova2006using}. This differs fundamentally from standard mixed-methods analysis, which often merges data concurrently to triangulate a single set of findings~\cite{creswell2025process}; instead, a sequential design serves as an overarching structural methodology in which integration occurs across phases~\cite{fetters2013achieving}. We started with an explanatory sequential mixed-methods design~\cite{kajamaa2020mixed} and, based on our observations and findings, moved to a multiphase sequential mixed-methods design~\cite{kajamaa2020mixed}. The theoretical foundation underpinning this approach is pragmatism, which justifies the use of complementary methods based on what most effectively addresses the evolving research problem~\cite{johnson2004mixed,shan2021philosophical}. For design scholars, this methodology is highly valuable because it provides a systematic chain of evidence, ensuring that technological interventions are directly anchored in prior empirical discoveries rather than designed in a silo. In our study, we applied this sequence to move progressively from understanding practices to system design. First, an exploratory survey and interviews (RQ1) revealed how urban Indian users leverage personal health information (PHI), highlighting a specific unmet interest in tools for shared, collective care. This key outcome directly dictated the focus of our next phase: we conducted co-design workshops~\cite{10.1162/codesmarc} and stakeholder interviews (RQ2) specifically to explore the ecosystem dynamics surrounding these collective needs. Finally, the user-informed requirements elicited from these participatory activities dictated the features of our Figma prototype, which we subsequently evaluated. Thus, our sequence was not a rigid template but a problem-driven chain where the outputs of each step justified the ``why'' and ``how'' of the next. We started the study in January 2024, following approval from the Institute's IRB. The co-design and prototype evaluation functioned as a targeted, theory-informed design probe used to (a) generate design concepts grounded in participants' existing practices and (b) collect usability and preference feedback to refine those concepts. The workflow of the overall study is presented in Figure~\ref{Figure 1}.

\subsection{Participants}
The study recruited participants from across urban India to capture a diverse range of ages, genders, and professional backgrounds. To capture variation among urban participants with access to PHI tools such as wearables, we focused on individuals from middle- to upper-middle socioeconomic strata, categorized using the updated Kuppuswamy scale (the latest peer-reviewed SES scale widely used in India for standardized adaptation inclusive of inflation) by Saleem et al.~\cite{saleemmodifiedkuppuswamysocioeconomic2021}. We employed a combination of purposive and snowball sampling methods to identify and engage participants. The sample comprised a diverse range of professions, including IT professionals, healthcare workers, students, entrepreneurs, and retirees, reflecting the broad spectrum of urban Indian society. We made deliberate efforts to ensure that all participants in the Interviews and subsequent parts of the study did not have any severe physiological/psychological conditions. This approach enabled us to gather insights from participants with diverse care practices and varying levels of familiarity and engagement with current systems and related tools, such as wearable health technologies (wearables). Prior to participation, all individuals were provided with a clear explanation of the study's purpose, and informed consent was obtained. The variation in participants’ ages, genders, and professions provided a broad range of perspectives on everyday care practices, PHI adoption and use, and the factors shaping engagement with wearable health technologies. Demographic details for the surveys and interview participants are summarized in Table~\ref{Table 1}, offering a comprehensive overview of the sample's composition.

\begin{table*}[ht!]
\centering
\begin{adjustbox}{width=\textwidth}
\begin{tabular}{|l | c | c | c | c | c | c | c | c | c|}
\hline
Study Component & Total &\multicolumn{3}{c|}{Gender} & \multicolumn{2}{c}{Age} & \multicolumn{3}{|c|}{Education}\\
\hline
& & Male & Female & Prefer Not & Range & Mean & High & Bachelors & Advanced\\
& & & & to Say & & & School & & Degrees\\
\hline
PHI awareness survey (RQ1) & 87 & 50 & 36 & 1 & 18-83 & 32 & 6 & 31 & 50\\
Open-ended questions (RQ1) & 34 & 20 & 14 & 0 & 18-62 & 32 & 1 & 16 & 17\\
Interviews (RQ1) & 22 & 14 & 8 & 0 & 18-62 & 28 & 5 & 6 & 11\\
6 Co-design workshops (RQ2) & 12 & 6 & 6 & 0 & 27-40 & 33 & 0 & 8 & 4\\
8 Stakeholder interviews (RQ2) & 12 & 7 & 5 & 0 & 29-76 & 33 & 1 & 6 & 5\\
Sharing awareness survey (RQ2) & 114 & 44 & 70 & 0 & 19-75 & 34 & 3 & 57 & 54 \\
\hline
\end{tabular}
\end{adjustbox}
\caption{\label{Table 1}Participant demographics from various study components. \textit{Advanced degree includes graduate degrees.} A total of 6 co-design workshops (12 participants) and 8 stakeholder interview sessions (12 participants).}
\end{table*}

\subsection{Study Design}
\subsubsection{PHI Awareness Survey} 
We launched our survey using SurveyMonkey and targeted a diverse demographic of urban Indian participants. The survey included both closed-ended and open-ended questions to capture a broad range of data on the use of PHI and related tools, such as wearables. The survey comprised 29 questions: 8 on demographic information and 8 open-ended (6 for users and 2 for non-users). The closed-ended questions focused on collecting demographic information (age, gender, occupation), the type of wearables used for PHI, frequency of use, and specific health metrics tracked (such as steps, heart rate, and sleep patterns). The open-ended questions focused on aspects of trust and usage, such as \textit{(1)Please explain briefly if you have ever changed the health wearable you use and why. (2) What do you understand about the health data/analysis from your wearable? (3) Do you trust the accuracy of the wearable health data? (4) Has the history from your health wearable ever helped you with doctor visits?} The detailed survey questionnaire is presented in Appendix~\ref{appendix:survey_rq1}. The open-ended survey questions allowed participants to elaborate on their experiences with wearables as part of their PHI, providing deeper insights into their motivations, challenges, and perceptions of these devices' utility for everyday proactive care. These responses were crucial for identifying recurring themes and patterns in the qualitative analysis. The survey remained open until response saturation was reached.

This survey received 99 responses, of which 87 were complete and used for further analysis. The demographics consisted of a mean age of 32 and a maximum age of 83, with 50 males, 36 females, and 1 participant who preferred not to disclose their gender. Among the complete responders, 55 users and 32 non-users were included. The majority (85\%) of the survey participants were between 18 and 44 years old.

\begin{table*}[ht!]
    \centering
    \begin{adjustbox}{width=\textwidth}
    \begin{tabular}{| l | l | l | l | l | l | l | l |}
    \hline
        \textbf{ID} & \textbf{Age} & \textbf{Gender} & \textbf{Education} & \textbf{Usage Purpose} & \textbf{Gifted} & \textbf{Usage Sustenance} & \textbf{Analytics Used} \\ \hline
        P1 & 19 & Male & High School & Health & Yes & Still Using & Advanced \\ 
        P2 & 25 & Female & Advanced Degree & Non-User & N/A & N/A & N/A \\ 
        P3 & 33 & Female & Advanced Degree & Health & Yes & Still Using & Advanced \\ 
        P4 & 33 & Female & Bachelors & Health/Utility & Yes & Stopped Using & Advanced \\ 
        P5 & 29 & Female & Bachelors & Utility & No & Still Using & N/A \\ 
        P6 & 20 & Male & High School & Fashion & No & Stopped Using & N/A \\ 
        P7 & 32 & Male & Doctorate & Health/Utility & No & Still Using & Basic \\ 
        P8 & 26 & Male & Advanced Degree & Health & No & Still Using & Basic \\ 
        P9 & 29 & Female & Bachelors & Health/Utility & Yes & Stopped Using & Advanced \\ 
        P10 & 62 & Male & Bachelors & Health/Utility & Yes & Still Using & Basic \\ 
        P11 & 34 & Male & Bachelors & Health/Utility & No & Stopped Using & Advanced \\ 
        P12 & 20 & Male & High School & Fashion & No & Stopped Using & N/A \\ 
        P13 & 29 & Male & Advanced Degree & Health/Utility & Yes & Still Using & Advanced \\ 
        P14 & 20 & Male & High School & Non-User & N/A & N/A & N/A \\ 
        P15 & 25 & Female & Advanced Degree & Fashion/Health & Yes & Still Using* & Basic \\ 
        P16 & 26 & Female & Advanced Degree & Health/Fashion & No & Still Using & Basic \\ 
        P17 & 26 & Male & Advanced Degree & Health/Utility & Yes & Still Using & Basic \\ 
        P18 & 18 & Male & High School & Health/Utility & No & Still Using & Advanced \\ 
        P19 & 36 & Male & Advanced Degree & Health/Utility & No & Stopped Using** & Advanced \\ 
        P20 & 29 & Male & Advanced Degree & Non-User & N/A & N/A & N/A \\ 
        P21 & 27 & Male & Bachelors & Health/Utility & Yes & Still Using & Basic \\ 
        P22 & 27 & Female & Advanced Degree & Health/Utility & Yes & Stopped Using & Basic \\
        \hline
    \end{tabular}
    \end{adjustbox}
    \caption{\label{Table 2} Interview participant demographics. \textit{* - Participant uses wearable rarely due to personal preference.}\textit{** - Participant stopped using wearables to wear normal watches and is looking for a reliable band/ring type health wearable.} \textit{Advanced degree includes graduate degrees.}}
\end{table*}

\subsubsection{Interview Design} 
To obtain richer qualitative insights beyond the survey, we conducted semi-structured interviews with 22 participants, all of whom were selected from the survey pool. The interviews were conducted based on participants' availability and continued until response saturation was reached regarding the emerging themes. Participants were selected to reflect on their everyday care practices and their experiences with supportive wearables. The interview group consisted of 14 males and 8 females, with a mean age of 28, representing diverse educational qualifications at the time of the interview, including up to high school, bachelor's degrees, and advanced degrees. Except for two participants (P16 and P22), who had PCOD but were otherwise healthy, the remaining participants had no chronic conditions. The detailed demographics are presented in Table~\ref{Table 2}. The interview group included 12 participants who actively tracked their health using wearables; the rest had never used a wearable or had stopped using them. This diversity in the interview sample enabled a comprehensive exploration of attitudes and behaviors towards proactive everyday care and the wearables that support it.

The interviews explored participants' broader health management practices and how wearables, or PHI usage more broadly, for proactive care fit within this ``ecology of care.'' To that end, participants were asked about their day-to-day care practices, routines, and their general habits (if any) towards proactive care. We also sought to understand the other actors involved in their care, the relationship dynamics in their day-to-day care practices, and the factors that influence them. The remaining interview questions delved deeper into themes constructed from the survey data on their current use of collective care systems. We covered topics such as the influence of cultural practices on wearable adoption, perceptions of data accuracy, and usability challenges associated with wearables for proactive care. While the initial interview guide was limited to wearable users, our review of the survey responses prompted us to include non-users to provide a deeper examination of their perspective on everyday care. This provided us with vital information about the factors that prevented them from becoming users, some of which helped us understand how current systems can be better aligned to support their care practices. This approach enabled the researchers to capture rich, contextual data, providing insights that went beyond what could be obtained from the survey alone. The detailed interview guide is presented in Appendix~\ref{appendix:interview_rq1}.

\begin{table*}[ht!]
    \centering
    \begin{tabular}{| l | l | l | l | c |}
    \hline
        \textbf{ID} & \textbf{Age} & \textbf{Gender} & \textbf{Education} & \textbf{Child(ren)}\\ 
        \hline
        CP1 & 34 & Female & Advanced Degree & 0\\ 
        CP2 & 35 & Male & Bachelors & 0\\ 
        CP3 & 33 & Female & Advanced Degree & 0 \\ 
        CP4 & 32 & Male & Bachelors & 0 \\ 
        CP5 & 27 & Female & Bachelors & 0\\ 
        CP6 & 30 & Male & Bachelors & 0\\ 
        CP7 & 33 & Male & Advanced Degree & 1\\ 
        CP8 & 31 & Female & Advanced Degree & 1\\ 
        CP9 & 29 & Female & Bachelors & 1\\ 
        CP10 & 31 & Male & Bachelors & 1\\ 
        CP11 & 38 & Female & Bachelors & 2\\ 
        CP12 & 40 & Male & Bachelors & 2\\
        \hline
    \end{tabular}
    \caption{\label{Table 3} Participant demographics of card sorting activity.}
\end{table*}

\subsubsection{Co-design Workshops and Stakeholder Interviews}
\begin{figure*}[ht]
    \centering
     \includegraphics[width=.95\linewidth]{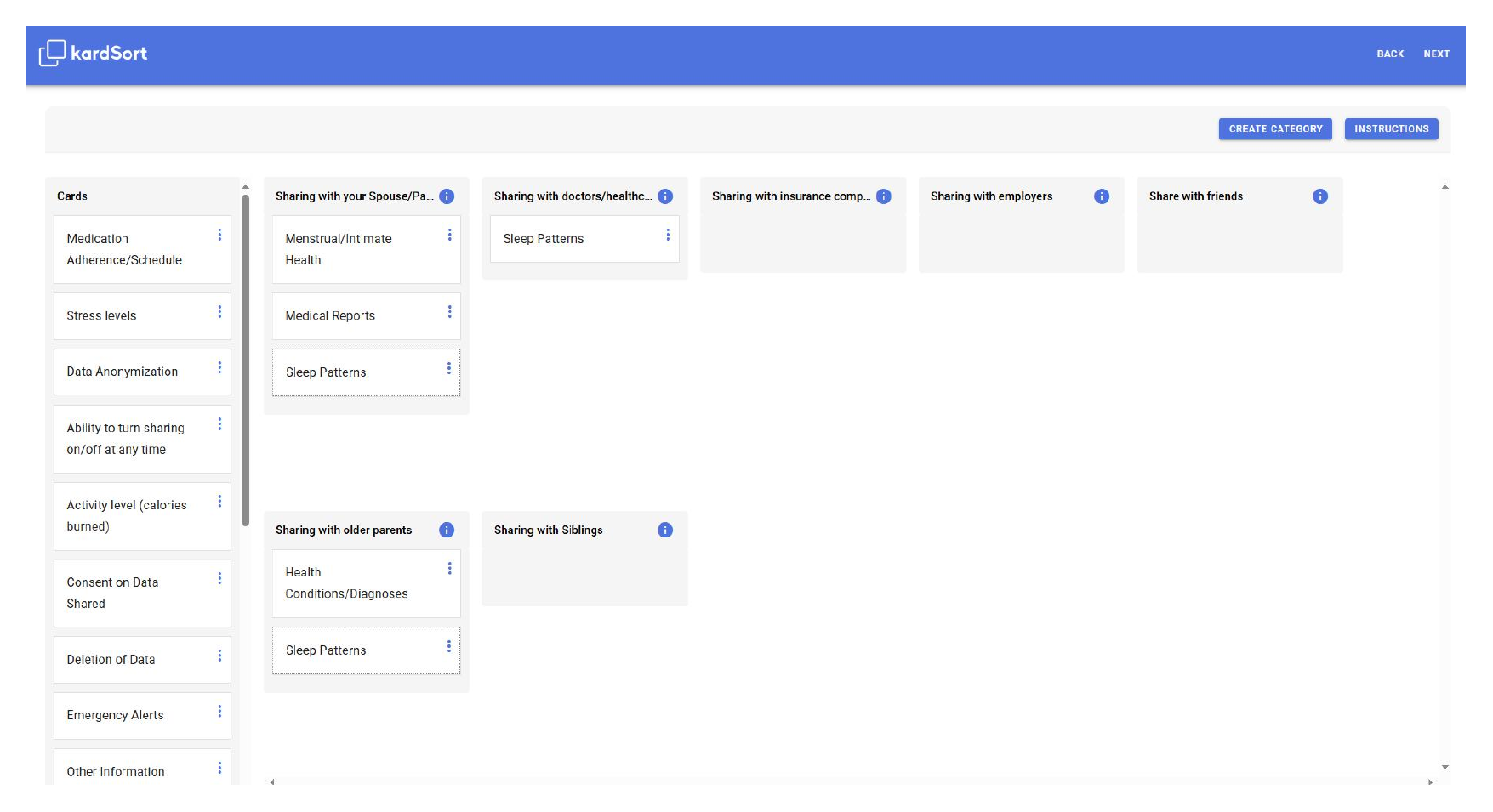}
     \caption{\label{Figure 2} Sample screenshot of card sorting activity.}
\end{figure*}

One of the critical findings from the analysis of survey and interview data was participants' desire for reliable tools to understand and share health information with various stakeholders for proactive care. However, most current health systems do not inherently provide such mechanisms, and those that do are ad hoc and require technological expertise. So, to understand \textit{who} the participants wanted to share with, \textit{how} and for how long they wanted to share it, and until \textit{when} they wanted to share it, we conducted co-design workshops with interviews for each Spouse/Couple pair (n=6) through dyadic card sorting activity using \textit{kartSort}. During the card-sorting activity, participant pairs (working individually while discussing with each other) moved data (such as vitals, GPS, etc.) and information cards (such as consent for storage, deletion, etc.) for different stakeholders, providing us with insights into their sharing preferences. A sample screen from the card-sorting activity is shown in Figure~\ref{Figure 2}. The different data/information cards could be duplicated, and the couple pairs could create new categories if needed for further granularity. The workshops were conducted until saturation to understand the variation in preferences among couples with and without a child (or children). Due to the cultural constraints of India, we were only able to evaluate these dynamics for cisgender heterosexual couples. The detailed demographics of the card sorting pairs are presented in Table~\ref{Table 3}. While these workshops provided insight into the dynamics of sharing among couples, we wanted to better understand other critical relationships, such as Parent-Child Relationships, and to gather input from doctors on the health information they believe is useful for supporting their patients. However, we skipped the workshops for them as older adults, and the doctors preferred interviews. This led us to conduct further stakeholder interviews with older parent-adult child pairs (M 64, F 34; M 76, M 40; F 72, F 33; F 75, F 50) and doctors (M 38, M 29, M 36, M 52). Details of the stakeholder interview guide are presented in Appendix~\ref{appendix:interview_rq2}.

\begin{figure*}[ht]
    \centering
     \includegraphics[width=.95\linewidth]{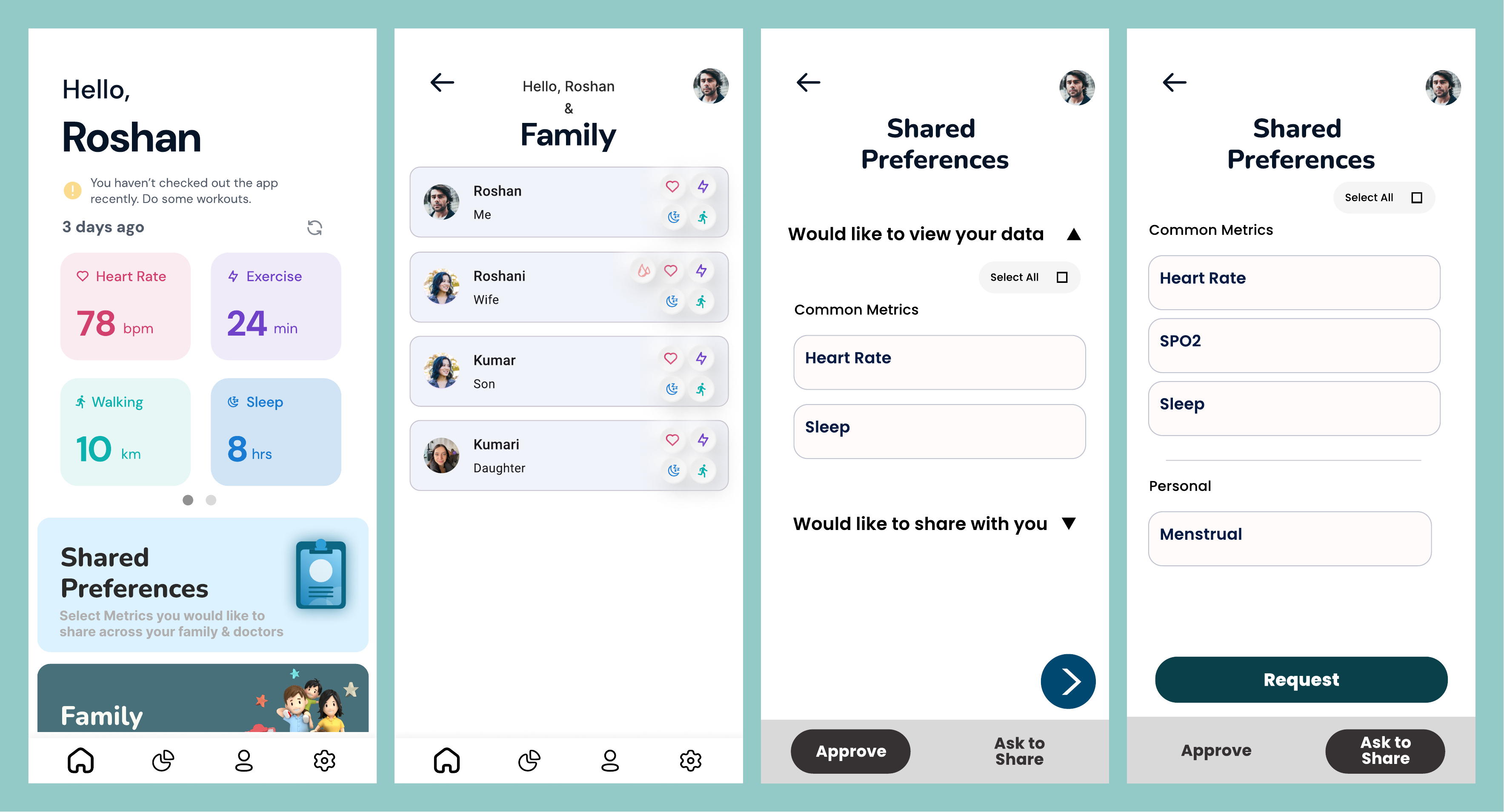}
     \caption{\label{Figure 3} Sample screenshots of the rapid figma prototype.}
\end{figure*}

\subsubsection{Design Probe and Sharing Awareness Survey}
After the co-design workshops and stakeholder interviews, we developed a medium-fidelity prototype in Figma. The interface was modeled using simpler, card-based formats with explicit consent requests to reflect participants' choices. We kept the interface simple and tile-based, with accessible screens, modular UI permissions, approval flows, and role templates. The prototype was iteratively designed and refined with participant feedback (2 iterations) before we finally evaluated its usability with participants across the study's components (n=38). This process helped us understand individuals' ideal sharing preferences, involving various stakeholders. Figure~\ref{Figure 3} gives a sample screenshot of the prototype. The figma prototype was initially evaluated by sharing it with the participants from co-design workshops for eliciting a qualitative evaluation, and further among participants from all components of the study with System Usability Scale (SUS) scores. The unanimous positive takeaway from the participants' feedback included its simplicity, modularity, and ease of control. Apart from these, the participant appreciated individual features and provided further recommendations, which are discussed later in the findings section. Due to repeated findings emerging around sharing, participants were questioned about existing mechanisms for sharing. However, many did not know these features existed, and those who did were reluctant to use them due to their technical complexity.

To gain a clearer picture of the current state of perception, awareness, and use of health information-sharing features, we designed a survey to assess these aspects among Android and iOS users. It comprised 26 items in total: five questions covering participant background and smartphone platform, followed by questions about health app and sharing feature use. Participants first reported their use of a health application (Apple Health for iOS users and health applications such as Google Fit, Fitbit, or Samsung Health for Android users) and then their use of its sharing feature. Based on their use of the sharing feature, participants answered questions addressing either their awareness, understanding, and use of the feature or their understanding, motivation, and interest in sharing health information. The non-user group included both participants who were aware of the sharing feature but did not use it and those who were unaware of it. Open-ended questions invited respondents to elaborate on their responses. The detailed survey questionnaire is presented in Appendix~\ref{appendix:survey_rq2}. The survey was closed once the desired number of responses was reached.

\begin{table*}[ht!]
\centering
\begin{adjustbox}{width=\textwidth}
\begin{tabular}{| l  |  c | c c | c c  |  c c | c c |}
\hline
& \multicolumn{5}{c|}{Health App Users} & \multicolumn{4}{c|}{Health App Non-Users}\\
\hline
Ecosystem & Users & Non-Users & SI & Unaware & SI & Aware & SI & Unaware & SI\\
\hline
Android & 8/65 & 35/65 & 31/35 & 22/65 & 20/22 & 6/34 & 5/6 & 28/34 & 26/28\\
iOS & 4/10 & 5/10 & 3/5 & 1/10 & 1/1 & 2/5 & 2/2 & 3/5 & 1/3\\
\hline
Total & 12/75 & 40/75 & 34/40 & 23/75 & 21/23 & 8/39 & 7/8 & 31/39 & 27/31\\
\hline
\end{tabular}
\end{adjustbox}
\caption{\label{Table 6}Participant counts across the sharing awareness survey (114 participants: 99 Android, 15 iOS) for health app users (75) and non-users (39). ``Users'' and ``Non-Users'' refer to use of the health-sharing feature, while ``Aware'' and ``Unaware'' indicate awareness of such a feature. ``SI'' reports the number interested in sharing health information relative to the immediately preceding group.}
\end{table*}

This survey (detailed in Table~\ref{Table 6}) was completed by 114 participants, with a mean age of 34 and a maximum age of 75, including 43 males, 70 females, and 1 who preferred not to say. Among the respondents (15 iOS/99 Android), only 12 (4 iOS/8 Android) used the health-sharing features; of the 102 participants who did not currently use a health-sharing feature (11 iOS/91 Android), 54 (4 iOS/50 Android) did not even know such a feature existed. This final survey helped reinforce the findings from the interviews and co-design workshops regarding the lack of suitable sharing mechanisms (or awareness of such features).

\subsection{Data Analysis \& Triangulation}

We analyzed the quantitative and qualitative data using analytic procedures appropriate to each data type and subsequently integrated findings across the study phases~\cite{fetters2013achieving,creswell2025process}. Quantitative survey responses were analyzed using descriptive statistics (frequencies and percentages) to characterize PHI adoption patterns and trust levels, while cross-tabulations examined associations between demographics and usage. For qualitative analysis, we applied Braun and Clarke's thematic analysis~\cite{braunUsingThematicAnalysis2006} to interview transcripts, open-ended survey responses, field notes, and co-design data. The authors collaboratively interpreted patterns and tensions across participants' accounts of everyday collective care practices and PHI use in those practices. Codes and sub-themes were inductively formulated around the tensions identified in the data. Through an iterative process of immersion, coding, and refinement, two principal themes were reflexively constructed: (1) \textit{Barriers to Proactivity} and (2) \textit{Enablers of Proactivity: Behaviors \& Triggers}. These themes characterize how PHI is perceived in the urban Indian context in relation to proactive collective care. The complete coding chart, including definitions, evidence sources, and design implications, is detailed in Table~\ref{Table 5} (Appendix~\ref{appendix:coding}).

Integration across the quantitative and qualitative components occurred by relating patterns identified in the survey to the qualitative accounts and by examining how findings from earlier study phases informed subsequent inquiry and design activities~\cite{fetters2013achieving,creswell2025process}. We additionally employed data triangulation~\cite{carterUseTriangulationQualitative2014} by examining findings across surveys, interviews, workshops, and stakeholder interviews, allowing us to identify areas of convergence and divergence across data sources. We then interpreted the resulting themes through the lens of Bronfenbrenner's Ecological Systems Theory (EST)~\cite{bronfenbrennerdevelopmentalresearchpublic1974}. This analysis showed that the barriers and enablers of proactive collective care were distributed across different ecosystem levels rather than being confined to the individual. Accordingly, the findings are organized across the different ecosystems of collective care, providing a theoretical grounding for understanding how factors that influence the use of health informatics technologies operate at interconnected levels of care.

\subsection{Positionality}
The authors of this study acknowledge their positionality as researchers from the Indian cultural context. While some authors still reside in India, bringing an insider's perspective, others live abroad with families in India. This diverse positionality enabled a comprehensive analysis that considered both local nuances and broader global trends in both proactive care practices and wearable use. The authors also recognize that their interpretations are influenced by their own experiences and backgrounds, which may shape the analysis and conclusions drawn. Throughout the research process, efforts were made to remain reflexive and mindful of potential biases, ensuring that the participants' voices were accurately represented and the analysis remained grounded in the data.
\section{Findings}
We present our findings across the ecosystems of collective care inspired by Urie Bronfenbrenner’s EST~\cite{bronfenbrennerdevelopmentalresearchpublic1974} to characterize the multilayered influences on participants' proactive collective care practices. Our analysis reveals that while participants value PHI for its proactive potential, its utility is shaped by a complex interplay of individual capability, relational dynamics, systemic infrastructure, and sociocultural practices. By organizing these observations into Micro, Meso, Exo, and Macro levels, we shift the unit of analysis from the individual to the collective care circle.

\subsection{Micro-system: The Negotiated Self between Embodied Intuition and Digital Signals}
The Micro level represents an individual's immediate internal capacities and relational foundations, driven by personal agency, trust, and health literacy. At the center of this ecosystem, the individual serves as the primary data steward for their care circle. However, this ``island of agency'' is often a space of deep negotiation between what the body feels and what the sensor reports. For many participants, the wearable is not an objective authority but a supplementary voice that must compete with a lifetime of ``embodied cues''\textemdash the internal sense that ``my body will let me know if something is wrong'' (P20), mirroring observations from Rapp \& Tirassa~\cite{Rapp02112017}.

Rather than adhering to clinical norms, users construct personalized ``internalized baselines.'' For instance, P1 (M, 19) established a self-observed $\approx90-100$bpm threshold as a prompt for mental health support. This metric was not derived from medical literature but from a perceived pattern in his own aging body. While this reflects a high degree of individual agency, it creates a literacy gap: when the ``subject of care'' relies on intuitive metrics, the information they share with the care circle remains unanchored in clinical reality, making collaborative sensemaking challenging. Survey data reflect this tension: while 60\% ($55/87$) of respondents track health metrics via wearables, their agency is often constrained by low health literacy and reliance on external, authoritative validation.

This literacy gap constrains collaborative sensemaking because the metrics users share with their care circles may lack a common clinical interpretation. The problem is further complicated by ``Gifted Agency'': many participants (P3, P4, P13, P15, P17) received devices as gifts and wore them out of obligation (P17) or novelty (P13) rather than with proactive intent. Consequently, users may receive ``numbers without clear next steps,'' leaving interpretation to clinicians:

\begin{quote}
    ``I do not understand whether the heart rate is good for me or not, the tracker does not tell me that.'' — P10 (M, 62)
\end{quote}

Engagement is also constrained by sensor skepticism and a preference for tangible, embodied feedback. Participants frequently cross-verify readings across devices rather than treating the wearable as an authoritative source. P4 (F, 33), for instance, noted that she does ``not blindly believe'' her tracker and will ``check with another and see whether it's me or it's the sensor'' before escalating a concern. Similarly, P5 (F, 29) contrasted the ``silence'' of smartwatches with the physical feedback of a manual blood pressure cuff:

\begin{quote}
    ``Maybe we are just used to a different kind of devices. For example, growing up, for the blood pressure machine [makes hand motions of pumping a cuff] (laughs), that thing, like you see something happening. (—) I think it's one of the perceptive Issues. And also because I'm not used it a lot. I don't have the perception, and its that.'' — P5 (F, 29)
\end{quote}

This preference for tangible feedback means that digital metrics are often discounted when they do not align with ``body-feeling.'' Survey respondents similarly described confirming readings with other devices or relying on a ``ballpark understanding'' because they recognized that sensors are not perfectly accurate. Physical friction points, such as strap allergies (P22) or battery maintenance, further disrupt continuous data collection and can lead to device abandonment.

The central breakdown at the Micro level is therefore interpretive rather than purely informational. Unclear or mistrusted metrics can generate anxiety instead of actionable clarity, as seen even with the high-literacy P11 (M, 34), who stopped tracking cholesterol metrics to relieve stress. When monitoring becomes burdensome or fails to produce interpretable action, individuals disengage from the longitudinal data stream, weakening the foundation for collective care.

\subsection{Meso-system: Relational Scaffolding and the Friction of Shared Sensemaking}
The Meso level encompasses direct care networks\textemdash including family, friends, healthcare workers, and paid caregivers\textemdash and the collaborative sensemaking among them. The lived reality within the participants' households reveals that health informatics shifts from a personal tool to a relational artifact managed within a \textit{care circle} of family members and trusted companions, woven into the ``quiet labor'' of familial duty. Tracking is a socially scaffolded practice where younger family members serve as primary conduits for both device adoption and data mediation. Extending the observations by Katule et al.~\cite{10.1145/2909609.2909664}, they frequently serve as the informational architects for older adults, longitudinally curating their health vitals from a distance to bridge the gap between clinical complexity and parental habituation. For instance, P8 described a practice of ``distributed vigilance,'' purchasing and configuring Fitbits for his parents to ``keep an eye'' on their metrics. While this repurposes devices for intergenerational safety, sharing remains a manual, human-driven process rather than a system-supported one.

Despite a strong latent interest in collaborative health management ($89/102$ sharing-feature non-users expressing interest in data sharing), existing technical abstractions often fail to support the dynamic nature of collective care. This is particularly salient in the Android ecosystem, where 48 of 99 survey participants had never heard of a built-in sharing mechanism, and another 41 found them too technical to use. When official platforms offer no clear pathway for collaboration, care circles instead rely on ad hoc channels such as a ``WhatsApp care infrastructure,'' manually moving health information between family members. Participants CP7 (M, 33) and CP8 (F, 31) described using messaging apps because dedicated sharing platforms were perceived as ``too technical'' or nonexistent in the Android ecosystem.

\begin{quote}
    ``We currently use whatever means possible (mostly just WhatsApp each other on health documents and information), we are open to a dedicated platform... '' — CP7 (M, 33) \& CP8 (F, 31)
\end{quote}

This workaround fragments the care record, leaving critical health history ``scattered all over the unsearchable phone storage'' (F, 33) and difficult to retrieve during clinical encounters or sudden health crises.

Simultaneously, collaborative sensemaking serves as social calibration: participants benchmark signals against peers, or ``confirm it with other devices of my friends,'' to anchor trust in uncertain metrics, as noted in survey responses. This shows that the care circle provides a necessary ``social anchor'' for data that might otherwise be dismissed as unreliable.

However, this shared agency is not without its relational costs. While couples like CP7 and CP8 favored open sharing, others viewed the care circle as a space for ``negotiated agency,'' where control over specific metrics remains paramount. Our prototype evaluation (Mean SUS = 78.16) further validated the preference for relational agency through modular sharing controls and ``ask to share'' approval flows. Participants favored role-based templates (e.g., ``weekly summary to family; alerts to clinician only'') to balance transparency and privacy. This need for granular control was echoed by CP6 (M, 30) and CP5 (F, 27), who drew on their experience with management-centric health applications in the UK to argue for higher-level abstractions in data disclosure:

\begin{quote}
    ``Sharing within spouses is OK, but what about the other stakeholders like employers or insurance companies? Do we get a say on what data goes to them or how they use it?... That level of control would be very useful.'' — CP6 (M, 30) \& CP5 (F, 27) 
\end{quote}

The Meso-system, therefore, reveals a persistent gap between the collective practices participants already enact and the limited support provided by existing systems. Care circles compensate for this gap through ``relational grit'': they manually coordinate, interpret, and selectively share health information, while negotiated control over who receives which data remains unresolved. The result is that collaborative sensemaking depends heavily on human effort rather than being reliably supported by the systems themselves.

\subsection{Exo-system: Goals Mismatch between Practice and Infrastructure}
The Exo level captures external structural environments, such as institutional care infrastructures, fragmented technologies, and mismatches between formal and informal care. The transition from familial care circles to professional healthcare represents a significant ``infrastructural cliff.'' While participants are often motivated by the proactive potential of their devices, this data frequently encounters an ``Institutional Gatekeeping'' that renders it medically invisible. This creates a profound mismatch between daily informal vitals and formal clinical requirements, which is further compounded by manufacturer silos. Ultimately, the resulting fragmentation across non-interoperable vendor applications increases cognitive load and erodes the value of longitudinal monitoring.

The most visceral evidence of this mismatch lies in the ``Clinic Gatekeeper'' phenomenon, where participant attempts to share data with clinicians and are met with rejection. Participant 11 (M, 34) recounted a pivotal moment when he presented his $SpO_{2}$ and heart rate data to a physician during a fever, only to have the professional ``shoo away'' the watch as a ``BS instrument.'' This institutional dismissal creates a ``Trust Vacuum.'' When formal systems reject patient-generated data, individuals stop viewing their health tracking as a clinical resource, as evidenced by survey responses indicating that metrics have ``never been relevant to any doctor visits'' or that they ``need not take help from the device'' because they rely solely on traditional check-ups.

Beyond the clinic door, the exo-system is further fragmented by ``Brand Sovereignty'' and the ``Compatibility Tax.'' Participant 3 (F, 33) highlighted frustration with technological silos, in which advanced health features such as ECG and Blood Pressure monitoring were locked behind the requirement to own specific smartphone brands, forcing care circles to choose between financial feasibility and functional utility. She characterized this as an ``expensive affair.''

\begin{quote}
    ``It's kind of an investment, like it is at least for me. So buying a watch which will cost 20,000/25,000 at the time of the release, it is kind of an expensive affair. So you cannot change these things with every update.'' — P3 (F, 33)
\end{quote}

This proprietary lock-in means that hardware updates disrupt continuity of care. Participant 13 (M, 29) described scouring the internet for discontinued APK files just to keep his older device functional after a phone upgrade, illustrating the ``informational precarity'' faced by users whose care is tethered to the volatile lifecycles of consumer electronics. Furthermore, this is complicated by the effort required to maintain a coherent health record across these silos, which was ``too much of a task,'' resulting in fragmented care histories that could not be easily used in clinical settings. P9 (F, 29) described the exhaustion of managing this fragmentation, noting that ensuring ``things are getting synced up'' was a significant burden that eventually led to disengagement:

\begin{quote}
    ``What I realized is that this was too much of a task for me to kind of, you know, (—) so that things are getting synced up.'' — P9 (F, 29)
\end{quote}

This friction is exacerbated by activity tracking algorithms that often misalign with daily household labor, which entails significant physical effort yet is rarely captured. P15 (F, 25) argued that the adoption of proactive health management would improve significantly if systems could integrate these traditional activities into their metrics:

\begin{quote}
    ``If you think about it, our lifestyles have been completely different from our parents. These trackers only work for us. If they are able to integrate regular daily activities (like cleaning the house, washing clothes, gardening) as part of measuring other metrics for health markers, I believe it can improve the adoption and create a more proactive health management mindset.'' — P15 (F, 25)
\end{quote}

Taken together, these infrastructural barriers leave patient-generated health data in an ``informational cul-de-sac''\textemdash capturing rich longitudinal vitals without a reliable pathway to clinical action. The resulting gap limits the clinical value of sustained tracking.

\subsection{Macro-system: The Cultural Logic of Resilience and Spiritual Assurance}
The Macro level reflects overarching societal influences, including sociocultural norms, the cultural logic of care, and shifts along a proactive-reactive axis. Beyond individual agency and relational dynamics, proactive care is governed by a broader cultural logic that often positions digital quantification as a secondary, or even intrusive, authority. We observed a hierarchy of assurance in which health is perceived as a state of spiritual and holistic balance rather than a series of data points, dictating not only how technology is used but also which forms of labor are recognized as ``care.''

For many participants, proactive health management is situated within spiritual and holistic frameworks in which digital tools are secondary to established forms of care. Participant 10 (M, 62), for example, used his smartwatch to monitor health parameters but located his fundamental sense of safety in ``God’s grace and my \textit{Guruji}’s (spiritual teacher) blessings.'' In this ``Spiritual-Digital Paradox,'' the device functions as a verification tool rather than a primary source of authority. A similar hierarchy appeared among younger participants: Participant 2 (F, 25) argued that ``if you are eating some good food, some healthy life will work better,'' framing the wearable as an ``extra thing'' that complicates a more familiar, holistic understanding of health. Across these accounts, care is associated with enduring practices of faith, diet, and traditional living, while the added work of technological tracking is perceived as supplementary rather than essential.

The flow of health information within the collective is further constricted by sociocultural norms regarding resilience and the performance of ``strength''. We identified a significant shift where individuals curate technology specifically to escape the ``burden of social judgment.'' Participant 15 (F, 25) explicitly preferred an AI health assistant:

\begin{quote}
    ``To be honest, sometimes I don't want to express how I feel [even when I am sad or down] to my parents... or someone... I don’t want to [show] my weakness to my friends or anyone in that way... I will think anyone/everyone can judge me.'' — P15 (F, 25) 
\end{quote}

In a culture where individual struggles are often subsumed by family stability, health informatics is repurposed as a ``Digital Sanctuary''\textemdash a private space for vulnerability that current social roles do not permit. Here, the logic of care is one of protection, not of the body, but of one's status within the care circle. 

Furthermore, the macro-system influences health through the lens of collective responsibility and familial obligation. In this context, health is a duty performed to ensure the continuity of the care circle. This collective orientation means that ``social motivators'' (including the culturally prevalent practice of family-led prompting) are often more effective triggers than automated device notifications. Participants in the stakeholder interviews and co-design sessions expressed a desire for features that leverage existing social pressures through communal rewards and visibility. CP1 (F, 34) highlighted how gamifying these cultural dynamics could foster a more engaged and proactive environment:

\begin{quote}
    ``Some family tracking and reward like a badge to the member who completes all goals, just to have encouraging competition in the family for better health would be nice.'' — CP1 (F, 34)
\end{quote}

At the Macro level, participants assigned meaning to health information through collective responsibility, spiritual assurance, and culturally familiar forms of care rather than individual optimization. Care was consequently situated within existing family and social practices, while the individualistic logic of many current systems remained misaligned with these expectations.

\subsection{Chrono-system: Temporal Shifts and Care Sustainability}
The Chrono level represents the temporal dimension, tracking how individual care needs, relationships, and system dynamics evolve over time. At the chrono-level, the temporal rhythms of life stages and digital habit obsolescence among individuals in a care circle directly impact the sustainability of collective care. There is a distinct transition toward proactive monitoring, accelerated by the ``pandemic pulse'' that ``being healthier is everyone’s priority'' (P9). This shift is reflected in the emergence of routine behaviors, such as goal-driven step counting and sleep tracking, which 60\% of our survey respondents reported as a baseline for daily wellness. For participants like P16 (F, 26), proactive engagement is driven by a long-term awareness of lifestyle-induced risks, such as those associated with ``desk jobs,'' leading to a consistent check on metrics:

\begin{quote}
    ``Since I have a desk job... I have to make sure that I'm doing at least, like so far that I come across the Internet, they do mention that you should be at least walking 5 kilometers if you are sitting all day and working. So that is one thing I keep in mind.'' — P16 (F, 26)
\end{quote}

Tool adoption follows a distinct temporal arc tied to evolving life stages. For younger participants, the device often enters their lives as a symbol of status or a ``cool'' trend before being discarded in favor of more professional markers of adulthood and its obligations as observed by Participant 6 (M, 20). Conversely, as participants age into their ``60s,'' the device shifts from a \textit{want} to a biological necessity. Participant 10 (M, 62) explicitly linked his adoption to this temporal milestone: ``I thought I'm turning to 60. So it is ideal to watch some parameters.'' This shift illustrates how the ``moral duty'' of care intensifies over time.

However, sustainability is frequently undermined by \textit{technological fatigue}. Our analysis identified a recurring ``honeymoon phase'' of tracking, typically lasting only 2 to 3 months, before maintenance outweighs the perceived health benefits. Participant 9 (F, 29) described this erosion, noting that the ``pain of charging the device every day'' eventually made it a burden rather than an enabler. Survey respondents echoed this, reporting they ``stopped using the Fitbit regularly'' because charging it alongside their ``phone, laptop, tablet, and AirPods'' became an unsustainable ``added effort.'' In these cases, the enactment of proactive care thus gives way to convenience, with users reverting to the ``good enough'' passive data collected by their smartphones.

Proactive care also requires longitudinal continuity, yet participants described ``informational precarity'' when transitions between devices or brands disrupted data. P19 (M, 36), for example, described the difficulty of maintaining a continuous record across models:

\begin{quote}
    ``I have been tracking for a long time... but then the watch changed, and getting things to sync again or maintaining that history across different models is difficult. You lose that continuity that you need for long-term health.'' — P19 (M, 36)
\end{quote}

Rather than introducing a new interoperability problem, this temporal disruption shows how the infrastructural barriers identified at the Exo level accumulate over time, turning longitudinal monitoring into episodic verification.

These experiences cause monitoring to oscillate between routine engagement and episodic use triggered by specific health concerns. Although participants' data indicated interest in sharing health and wellbeing information, practice often remained ``short-term and goal-oriented,'' with deeper collective analysis and clinician contact reserved for symptomatic episodes. The Chrono-system, therefore, reveals a central sustainability problem: the desire for proactive collective care may be strong, but maintaining it requires longitudinal continuity across changing habits, life stages, and consumer technologies. Across the ecological levels, proactive collective care emerges as a practice constrained not by a single barrier but by interacting tensions in interpretation, relationships, infrastructure, culture, and time.

\subsection{Key Observations from the Study}
Taken together, the five ecological levels reveal five overarching observations that characterize proactive collective care within our study, serving as analytical anchors for our discussion. 
\textbf{(1) Relational Interpretation:} Health information is interpreted relationally rather than solely by the individual, with embodied knowledge, personal health literacy, and trusted others shaping how digital signals are understood. 
\textbf{(2) Unsupported Coordination:} Collective care already entails substantial human coordination, but existing PHI systems offer limited support for sharing, negotiating, and maintaining health information across care circles. 
\textbf{(3) Trust and Uncertainty:} Trust and uncertainty strongly influence whether health signals become actionable, as participants frequently relied on cross-verification or familiar social anchors when digital metrics were ambiguous. 
\textbf{(4) Systemic Fragmentation:} Technological and institutional fragmentation disrupts the transition from monitoring to coordinated action, particularly when data cannot move coherently across devices, applications, and clinical settings. 
\textbf{(5) Temporal Dynamics:} Proactive collective care is temporally dynamic, meaning care needs, motivations, relationships, and technologies change over time, making sustained engagement dependent on longitudinal continuity and adaptation.
\section{Discussion and Design Implications}

The observations synthesized collectively point to a central tension: proactive collective care is not simply a matter of improving individual tracking, but of supporting interpretation, coordination, trust, infrastructure, and continuity across a care circle. While our participants expressed a desire for earlier intervention, our findings reveal that ``proactivity'' is frequently stalled by structural barriers\textemdash specifically ecosystem fragmentation, low digital health literacy, and a conditioned skepticism of ``black-box'' metrics. Prior CSCW and HCI literature on family informatics~\cite{bindaintergenerationalsharinghealth2017, chasharedresponsibilitycollaborative2024}, care networks~\cite{ugargolfamilycaregivingolder2018}, and interpersonal personal informatics~\cite{murnanepersonalinformaticsinterpersonal2018} has established that health tracking is socially situated and relational. However, existing frameworks largely model health information sharing as an \textit{output or extension} of an individual tracking loop. In contrast, by interpreting these friction points through the lens of EST~\cite{bronfenbrennerdevelopmentalresearchpublic1974} and Wozniak et al.’s concept of \textit{Health Information Ecologies}~\cite{wozniak-oconnorHealthInformationEcologies2024}, we argue that supporting sustainable adoption requires moving beyond the individual as the primary unit of analysis, reframing health systems as a distributed sociotechnical scaffold that supports the collective responsibility of care circles.

These observations also clarify the relationship between the two research questions. RQ1 examined how urban people engage in proactive collective care, revealing a dualism of individual motivations (goal-driven tracking similar to individual-centric observations~\cite{chmielewskiDiscussionsEndUsers2024}) and collective caregiving practices (family-mediated monitoring reflecting prior work on supporting older adults~\cite{lieberchangingfamilystructures2020, ugargolfamilycaregivingolder2018, gustafssoninformalcaregivingperspectives2022}, chronic conditions~\cite{bhatwearehalfdoctors2023}, and pediatric management~\cite{chasharedresponsibilitycollaborative2024}). Across these practices, uncertainty at the individual level can generate relational workarounds; those workarounds can encounter infrastructural barriers; and the resulting disruptions can reinforce reactive rather than proactive care. RQ2, therefore, asked what system properties could support collective care across these interconnected constraints. We synthesize these requirements into three operational capabilities: (a) \textit{Agency}, which supports granular and transparent control over what is shared and with whom; (b) \textit{Elicitation}, which converts uncertain or easily ignored signals into contextually appropriate opportunities for action; and (c) \textit{Engagement}, which bridges literacy, interpretive, and clinical barriers to translate health information into collective action. Together, these capabilities form the core of ``\textbf{CC-Proact}'', an operational design map that translates the five empirical observations into concrete interaction levers for proactive collective care.

\begin{figure*}[ht]
    \centering
     \includegraphics[width=.95\linewidth]{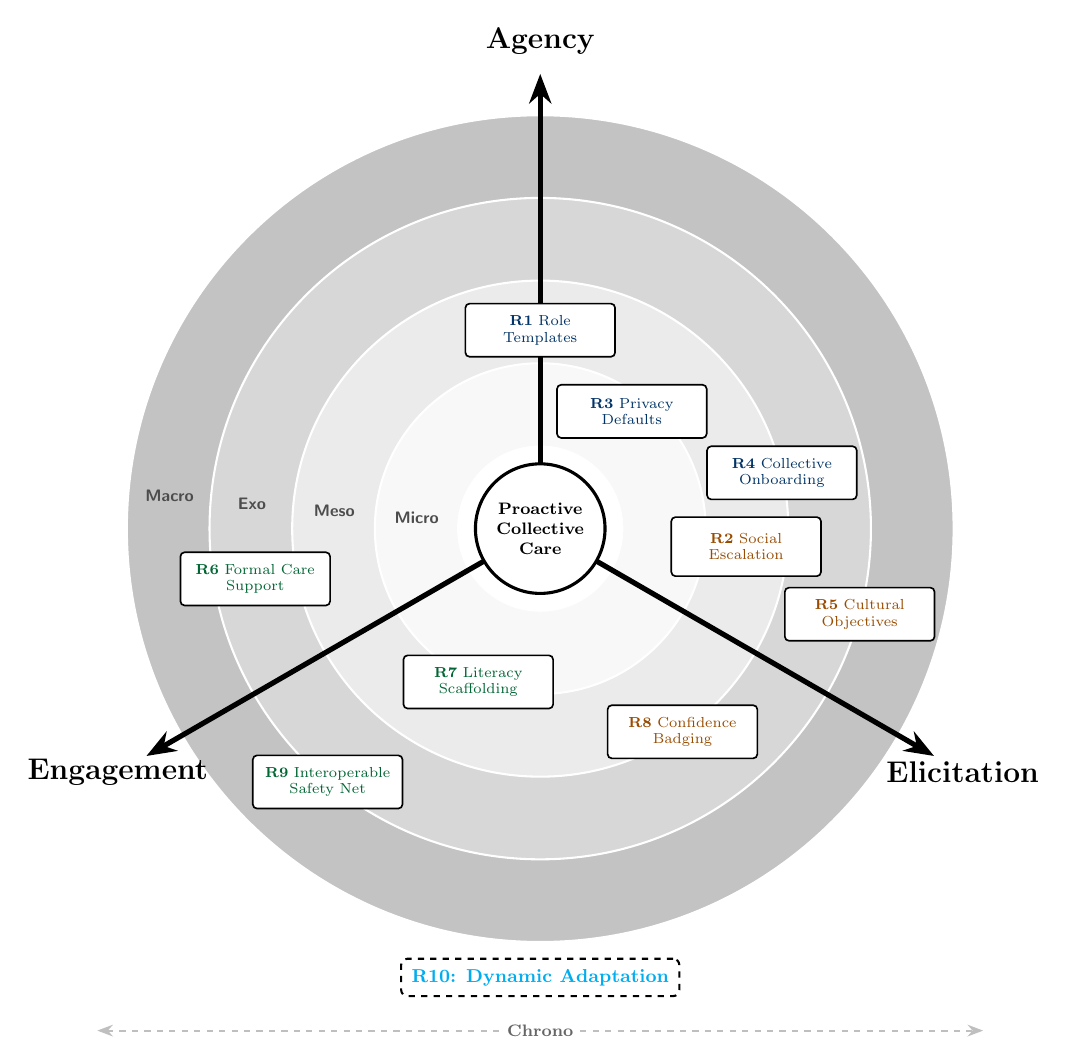}
        \caption{\label{Figure CC} CC-Proact Design Map. The map translates ecological influences on proactive collective care into three design capabilities: Agency, Elicitation, and Engagement, and maps six recommendations according to the ecological systems they address and the capabilities they emphasize.}
\end{figure*}

\subsection{CC-Proact: Agency, Elicitation, \& Engagement}

Drawing on the study observations and the ``Barriers to Proactivity'' and ``Enablers of Proactivity'' themes identified in our thematic analysis, CC-Proact operationalizes ecological insights into prescriptive design levers. Through CC-Proact, we explicitly specify how health technologies should function across complex sociotechnical webs by categorizing system requirements into three interdependent interaction domains: Agency (the relational who), Elicitation (the contextual when), and Engagement (the scaffolding how).

\begin{itemize}
    \item \textbf{Agency (The Relational ``Who''):} Grounded in our findings that device ownership rarely equates to sole data stewardship, participants frequently delegated monitoring to family members (``trusted proxies''), challenging the ideal of the autonomous ``quantified self.'' To support this, the agency has to move beyond binary access controls to operationalize what Kaziunas et al. term ``Data Care''~\cite{10.1145/2998181.2998303}\textemdash reframing data maintenance as a shared relational practice across micro- and meso-ecosystems.
    \item[] \textit{Design Implication:} Systems must implement ``Role Templates'' (e.g., \textit{Caregiver}, \textit{Clinician}, \textit{Observer}) that afford granular control over data types and temporal windows. This formalizes ad-hoc sharing workarounds while preserving individual autonomy and social safety.
    \item \textbf{Elicitation (The Contextual ``When''):} Derived from participant reports of notification fatigue and sensor distrust, Elicitation addresses the system's ability to surface signals intelligently across micro and exo barriers. Building on personal health design spaces~\cite{7445786}, Elicitation adapts alerts to contextual conditions and perceived reliability, consistent with prior work on adaptive and just-in-time intervention timing~\cite{10.1145/3613904.3642747}. Prior work has also shown how a user's immediate social context can be used to identify appropriate moments for notification delivery~\cite{10.1145/2998181.2998189}. Elicitation extends these contextual approaches to the relational context of the care circle, where social roles and shared responsibility influence not only when a prompt should be delivered, but also who may need to act on it.
    \item[] \textit{Design Implication:} Systems should integrate ``Confidence Badges'' to transparently communicate sensor uncertainty~\cite{10.2196/31618} and utilize ``Escalation Paths'' (e.g., nudging the user first, then escalating to a care circle proxy) that mirror established social support structures.
    \item \textbf{Engagement (The Scaffolding ``How''):} Grounded in observed literacy gaps and clinical gatekeeping, Engagement operationalizes the supportive layer of ecological health models~\cite{wozniak-oconnorHealthInformationEcologies2024}. It bridges the gap between raw data collection and actionable sensemaking for users with varying digital health literacy~\cite{leepatientstechnologyreadiness2022}.
    \item[] \textit{Design Implication:} Interfaces must prioritize plain-language ``Micro-Guidance'' and concise ``Clinician Summaries'' over dense data streams. This shifts the interpretive burden from the isolated individual to the collective care circle, or even to the broader healthcare provider.
\end{itemize}

As an operational design map (Figure~\ref{Figure CC}), CC-Proact centers on relational interdependence rather than personal mastery~\cite{lazarwhyweuse2015}. Grounded in Kirchner et al.’s translational design framework~\cite{10.1145/3411764.3445587}, CC-Proact embeds the realities of developing health infrastructures and hierarchical care structures directly into the system’s mechanics, offering a structured basis for designing collective-centered proactive systems.

\subsection{Recommendations for Designers and Developers}

Using our CC-Proact mapping, we derive six higher-level design recommendations (Table~\ref{Table 4}) that consolidate the design requirements identified across the study. These recommendations do not claim novelty for the underlying interaction mechanisms; rather, they synthesize and adapt established approaches to the relational, ecological, and longitudinal conditions of collective care. While data sharing, prompting, privacy controls, health-literacy support, and interoperability are established in prior HCI and health informatics research, CC-Proact contributes by configuring these capabilities for collective care rather than for ``individual'' users. The six recommendations operationalize CC-Proact across ecological layers, spanning Agency, Elicitation, and Engagement, while extending these capabilities across the Chrono-system.

\begin{longtable}{@{}p{0.02\textwidth} p{0.97\textwidth}@{}}
\caption{\label{Table 4} Recommendations for designing Proactive Collective Care systems mapped to CC-Proact capabilities.}\\

\toprule
\endfirsthead

\multicolumn{2}{@{}l}{\textit{Table~\ref{Table 4} (continued)}} \\
\toprule
\endhead

\midrule
\multicolumn{2}{r}{\textit{(continued on next page)}} \\
\endfoot

\bottomrule
\endlastfoot

\textbf{R1} &
\begin{tabular}[t]{@{}p{\linewidth}@{}}
\textbf{[Agency] Relational Agency \& Care-Circle Workflows:} \newline \textit{The Gap:} Standard PHI systems often treat sharing as a binary ``share/don't share'' decision and assume that onboarding, device setup, and data maintenance are individual activities, overlooking the roles, power dynamics, and support relationships within a care circle. \newline \textit{Recommendation:} In collective care, systems should treat agency as relational and support a ``collective workflow'' from the outset. Designs should provide role-based templates (e.g., \textit{Caregiver}, \textit{Observer}, \textit{Clinician}) with granular permissions for data types and temporal windows, while allowing users to establish a Care Circle and invite trusted members to support device setup, maintenance, and information management. This formalizes the ad hoc sharing and support practices observed in the study while preserving individual autonomy and negotiated control~\cite{10.1145/3240925.3240936, bindaintergenerationalsharinghealth2017, leefamilyscopevisualizingaffective2024}.
\end{tabular} \\

\midrule

\textbf{R2} &
\begin{tabular}[t]{@{}p{\linewidth}@{}}
\textbf{[Agency \& Elicitation] Socially Situated Elicitation:} \newline
\textit{The Gap:} Individual-centric notifications can be ignored in high-friction environments, while abstract health goals may fail to resonate with participants' lived routines and culturally meaningful forms of care. \newline
\textit{Recommendation:} Systems should support context-sensitive prompting that can escalate through trusted members of the care circle when action is not taken, while also allowing users to frame goals around locally meaningful activities and responsibilities. Configurable ``Escalation Paths'' should move from user-centric nudges to trusted proxies when appropriate, and goal templates should incorporate familiar activities such as household work, family responsibilities, and other culturally meaningful routines. Building on prior work on adaptive intervention timing~\cite{10.1145/3613904.3642747} and social-context-aware notification timing~\cite{10.1145/2998181.2998189}, this recommendation extends contextual elicitation from the individual user to the relational context of the care circle, where prompting is coupled with shared responsibility for subsequent action.
\end{tabular} \\

\midrule

\textbf{R3} &
\begin{tabular}[t]{@{}p{\linewidth}@{}}
\textbf{[Agency \& Elicitation] Privacy-Preserving Defaults \& Calibrated Trust:} \newline \textit{The Gap:} Consent fatigue, uncertainty about downstream data use, and skepticism toward sensor accuracy can undermine both safe sharing and continued reliance on health technologies. \newline \textit{Recommendation:} Systems should minimize unnecessary retention, use example-based consent to communicate potential social consequences of sharing, and make the provenance and uncertainty of derived health metrics visible. Privacy-preserving defaults (e.g., time-bounded retention) should be complemented by ``Provenance Tags'' and ``Confidence Badges'' that indicate how a metric was generated and how much uncertainty it carries. Together, these mechanisms allow users and care-circle members to make more informed disclosure decisions and appropriately calibrate their reliance on digital health signals~\cite{10.1145/3613904.3642815, 10.2196/31618}.
\end{tabular} \\

\midrule

\textbf{R4} &
\begin{tabular}[t]{@{}p{\linewidth}@{}}
\textbf{[Engagement \& Elicitation] Interpretation-First Health-Literacy Scaffolding:} \newline \textit{The Gap:} Raw health metrics can be opaque or anxiety-inducing, placing the burden of interpretation on individuals who may lack the digital or health literacy required to make sense of them. \newline \textit{Recommendation:} Interfaces should prioritize ``Micro-Guidance'' and progressive disclosure over dense streams of raw metrics. Primary views should provide plain-language explanations of what a signal may mean and what, if anything, a user might consider doing next, while allowing deeper detail to be accessed when needed. These explanations should be designed to support both individual understanding and collaborative sensemaking within the care circle, reducing reliance on ad hoc or potentially unreliable external sources for interpretation~\cite{10.1145/3301275.3302317}.
\end{tabular} \\

\midrule

\textbf{R5} &
\begin{tabular}[t]{@{}p{\linewidth}@{}}
\textbf{[Engagement] Clinical Integration \& Longitudinal Interoperability:} \newline \textit{The Gap:} Fragmented device ecosystems and limited clinical pathways can leave patient-generated health data isolated across applications, difficult to maintain across device changes, and difficult for clinicians to interpret or use during consultations. \newline \textit{Recommendation:} Systems should provide a ``last-mile'' pathway from everyday tracking to clinical action by generating concise, time-windowed ``Clinician Summaries'' and supporting interoperable, portable health records through standard formats such as FHIR or CSV. This combination enables longitudinal assembly of health information across devices and applications while reducing the burden on clinicians, helping preserve continuity when hardware, vendors, or applications change~\cite{mandl2020push}.
\end{tabular} \\


\textbf{R6} &
\begin{tabular}[t]{@{}p{\linewidth}@{}}
\textbf{[Chrono-system] Dynamic Adaptation Across Time:} \newline \textit{The Gap:} Static health goals and engagement measures do not account for changing life stages, evolving care relationships, seasonal or episodic health needs, or the obsolescence of consumer technologies, contributing to the ``honeymoon phase'' and eventual abandonment of tracking. \newline \textit{Recommendation:} Systems should support longitudinal adaptation by allowing individuals and care circles to change objectives as circumstances evolve, while evaluating proactivity through measures tied to meaningful action rather than superficial engagement. Metrics such as ``Alert-to-Action Latency'' or ``Clinician-Confirmed Detections'' can capture whether information contributes to timely action. By allowing goals to transition from general wellness to context-specific care needs, systems can remain useful as active care scaffolds across changing temporal conditions.
\end{tabular} \\

\end{longtable}

Grounded in empirical participant data and evaluated via our Figma prototype as a theory-informed design probe, these recommendations demonstrate design plausibility for collective health systems. Together, they translate the study observations into a compact set of design priorities through the three CC-Proact capabilities, rather than a collection of isolated features. In settings where family caregiving and systemic fragmentation are salient, this mapping demonstrates how ecological theory can directly inform design and evaluation to support proactive collective care, concretely answering our RQ2.

\subsection{Limitations and Future Work}

While our study offers novel insights into the personal, relational, and cultural dimensions of collective care practices, we acknowledge several limitations. First, our qualitative analysis draws on surveys, interviews, and co-design workshops; while these methods elicit rich empirical accounts, they rely on self-reported behaviors and intentions rather than longitudinal behavioral observation in the field. Second, our prototype evaluation served as a theory-informed design probe to assess perceived usability, preferences, and design plausibility, rather than as evidence of long-term real-world adoption or clinical impact. Validating sustained impacts on care coordination and clinical outcomes will require longitudinal field deployments of a fully functioning system. Finally, because generalizability across all socioeconomic strata was not an objective of this work, our sample focused primarily on urban, middle- and upper-middle-class participants in India with (potential) access to consumer digital health tools. Focusing on relatively affluent demographics introduces the risk of creating or compounding \textit{Intervention-Generated Inequality} (IGI)~\cite{10.1093/jamia/ocy052}. Informatics interventions designed around digitally literate, resource-rich care circles can inadvertently widen health disparities if they presume access to proprietary hardware, high digital literacy, or stable smartphone ecosystems. In lower socioeconomic contexts where digital wearables are absent and pen-and-paper or informal verbal tracking remain the primary modes of health management, different structural barriers exist. Future work should therefore examine how collective health informatics can be scaffolded across lower-SES settings\textemdash for example, through hybrid digital-paper interfaces or community health worker mediation\textemdash and should evaluate how collective-care systems perform amid changing care needs, relationships, and technological conditions over time.

Our ongoing research direction addresses these needs through the design and development of a system that supports general health and wellbeing as well as collective care practices in response to medical inquiries. The system is intended to assist users and their care circles in interpreting and contextualizing personal health data aggregated from diverse sources, presenting information in formats that are comprehensible and actionable for different roles. In particular, we plan to integrate clinical documents (e.g., physician reports, laboratory results, and medication records) alongside information from wearable health devices, using multimodal data synthesis to generate personalized, holistic insights (similar to the objective of Moore et al.'s work~\cite{10.1145/3494964}) for members of the care circle, to support their engagement and health awareness (akin to the observations of Lin et al.~\cite{10.1145/3678575}). The interface will employ a conversational, chatbot-style interaction to support intuitive engagement, mediate family sharing norms, and provide literacy-appropriate explanations. It will also provide access to evidence-based health, medical, and wellness content to address health misinformation and strengthen public health literacy. This planned work will allow us to examine whether the capabilities and recommendations proposed by CC-Proact can be translated into sustained, real-world support for collective care.
\section{Conclusion}

This study examined how people in urban India practice collective care and how existing PHI systems support or constrain proactive care within these relationships. Across our mixed-methods study, we found that proactive collective care is shaped by interacting tensions across individual interpretation, relational coordination, technological and clinical infrastructure, sociocultural expectations, and continuity over time. Rather than treating these tensions as isolated usability problems, our ecological analysis shows how they accumulate: uncertainty at the individual level can drive relational workarounds, fragmented workarounds can encounter infrastructural barriers, and these disruptions can reinforce reactive rather than proactive patterns of care.

These findings motivate CC-Proact, an operational design map that translates ecological influences into three capabilities\textemdash Agency, Elicitation, and Engagement\textemdash and structures ten design recommendations for collective care. The prototype evaluation provides evidence of participant preferences and design plausibility for coordinated, trustworthy sharing and sensemaking, while the study does not establish long-term adoption or clinical impact. Taken together, the work reframes proactive care from an individual-tracking problem to a collective, longitudinal practice and provides a structured basis for designing systems that support this reality.

\bibliographystyle{ACM-Reference-Format}
\bibliography{references}
\clearpage 





\appendix
\section{Thematic Analysis Coding Chart}
\label{appendix:coding} 

\begin{longtable}{@{}p{0.12\textwidth} p{0.51\textwidth} p{0.31\textwidth}@{}}
\caption{\label{Table 5} Coding chart (Using qualitative data from various study components).}\\

    \toprule
    \textbf{Themes} & \textbf{Sub-Theme/Code} & \textbf{Design implication} \\
    \midrule
    \endfirsthead
    
    \multicolumn{3}{@{}l}{\textit{Table~\ref{Table 5} (continued)}} \\
    \toprule
    \textbf{Themes} & \textbf{Sub-Theme/Code} & \textbf{Design implication} \\
    \midrule
    \endhead
    
    \midrule
    \multicolumn{3}{r}{\textit{(continued on next page)}} \\
    \endfoot
    
    \bottomrule
    \endlastfoot
    
    \textbf{Barriers to Proactivity} &
    \begin{tabular}[t]{@{}p{\linewidth}@{}}
        \textbf{Ecosystem Fragmentation \& Inter-op Problems}\\
        \textit{Definition:} Friction from multiple non-interoperable apps/brands and scattered records preventing longitudinal PHI use. \\
        \textit{Evidence (source):} Interviews.\\
    \end{tabular} &
    \begin{tabular}[t]{@{}p{\linewidth}@{}}
        Design interoperable dashboards, import/export (EHR) connectors, and a single-view aggregator; prioritize Android-first flows based on market share.\\
    \end{tabular} \\
    
    &
    \begin{tabular}[t]{@{}p{\linewidth}@{}}
        \textbf{Low Digital Health Literacy}\\
        \textit{Definition:} Users cannot interpret metrics or know what action to take from raw analytics. \\
        \textit{Evidence (source):} Field notes and Interviews.\\
    \end{tabular} &
    \begin{tabular}[t]{@{}p{\linewidth}@{}}
        Provide literacy-appropriate threshold-based plain-language recommendations and visual metaphors.\\
    \end{tabular} \\
    
    &
    \begin{tabular}[t]{@{}p{\linewidth}@{}}
        \textbf{Trust \& Sensor-Confidence Issues}\\
        \textit{Definition:} Doubts about the accuracy of advanced metrics; users cross-check multiple sources before acting. \\
        \textit{Evidence (source):} Interviews and Co-design workshops\\
    \end{tabular} &
    \begin{tabular}[t]{@{}p{\linewidth}@{}}
        Add sensor-confidence indicators, provenance badges, cross-device comparison views, and clinician-validated thresholds.\\
    \end{tabular} \\
    
    &
    \begin{tabular}[t]{@{}p{\linewidth}@{}}
        \textbf{Device \& Form-factor Friction}\\
        \textit{Definition:} Physical discomfort, charging burden, and undesired form factors cause abandonment. \\
        \textit{Evidence (source):} Interviews\\
    \end{tabular} &
    \begin{tabular}[t]{@{}p{\linewidth}@{}}
        Support multiple modalities (bands, rings, app-only), low-maintenance devices, battery-friendly modes, and non-wearable capture options.\\
    \end{tabular} \\
    
    &
    \begin{tabular}[t]{@{}p{\linewidth}@{}}
        \textbf{Preference for Embodied/Traditional Cues}\\ \& \textbf{Reactive Norms}\\
        \textit{Definition:} Reliance on body-feeling and doctor visits; belief that reactive care is sufficient. \\
        \textit{Evidence (source):} Interviews and Co-design workshops.\\
    \end{tabular} &
    \begin{tabular}[t]{@{}p{\linewidth}@{}}
        Design a hybrid onboarding that maps PHI outputs to familiar embodied cues; provide low-friction “check only when alerted” modes.\\
    \end{tabular} \\
    \midrule
    
    \textbf{Enablers of Proactivity: Behaviours \& Triggers} &
    \begin{tabular}[t]{@{}p{\linewidth}@{}}
        \textbf{Stakeholder specific sharing}\\
        \textit{Definition:} PHI used as a shared family resource; younger family members' recommendations create derived trust; Linking PHI to clinicians enables earlier action. \\
        \textit{Evidence (source):} Interviews and health-sharing survey: 114 respondents — 20 users, 54 non-users, 40 unaware.\\
    \end{tabular} &
    \begin{tabular}[t]{@{}p{\linewidth}@{}}
        Create a user-friendly family-sharing UI with approval/request features. Include one-page PDF/HL7-style exports, customizable clinician summaries, and time-based trend highlights.\\
    \end{tabular} \\
    
    &
    \begin{tabular}[t]{@{}p{\linewidth}@{}}
        \textbf{Motivational Social Features}\\
        \textit{Definition:} Family challenges, badges, and friendly competition that encourage preventive routines. \\
        \textit{Evidence (source):} Co-design workshops.\\    
    \end{tabular} &
    \begin{tabular}[t]{@{}p{\linewidth}@{}}
        Add consensual family challenges, non-punitive badges, group goals, and shareable progress snapshots that respect privacy controls.\\
    \end{tabular} \\

    &
    \begin{tabular}[t]{@{}p{\linewidth}@{}}
        \textbf{Actionable Explanations (literacy-appropriate)}\\
        \textit{Definition:} Plain-language recommendations tied to metric thresholds (what to do, not just numbers). \\
        \textit{Evidence (source):} Interviews and Co-design workshops.\\
    \end{tabular} &
    \begin{tabular}[t]{@{}p{\linewidth}@{}}
        Convert numbers $\rightarrow$ actions: “If X over Y days, do Z”; contextual links to vetted local resources; in-app “explain this metric” with examples.\\
    \end{tabular} \\
    
    &
    \begin{tabular}[t]{@{}p{\linewidth}@{}}
        \textbf{Proactive Monitoring (behavior)}\\
        \textit{Definition:} Regular, routine checking of PHI and prevention-oriented actions (daily/weekly review, goal setting, early consultations). \\
        \textit{Evidence (source):} 55/87 ($\approx$60\%) survey participants reported tracking with a wearable; interviews\\
    \end{tabular} &
    \begin{tabular}[t]{@{}p{\linewidth}@{}}
        Quantify proactive vs reactive segments; surface summarized trends (daily/weekly) and goal dashboards; surface family-monitoring affordances.\\
    \end{tabular} \\
    
    & 
    \begin{tabular}[t]{@{}p{\linewidth}@{}}
        \textbf{Triggers for Proactivity}\\
        \textit{Definition:} Specific events, alerts, family prompts or clinician requests that cause earlier action (alerts, family messages, prototype “ask to share”). \\
        \textit{Evidence (source):} Co-design workshops\\
    \end{tabular} &
    \begin{tabular}[t]{@{}p{\linewidth}@{}}
        Implement configurable alerts, “ask-to-share” nudges, permissioned push messages, and simple export/share buttons for clinician workflows.\\
    \end{tabular} \\
\end{longtable}

\section{Survey Questionnaire (RQ1)}
\label{appendix:survey_rq1}

\begin{tcolorbox}[enhanced,breakable,
                  colback=white,colframe=black,boxrule=0.6pt,arc=1pt,
                  left=6pt,right=6pt,top=6pt,bottom=6pt,
                  before skip=6pt,after skip=6pt]
\footnotesize
\setlist[itemize]{noitemsep, topsep=0pt, left=12pt}
\setlist[enumerate]{noitemsep, topsep=2pt, left=0pt, label=\textbf{Q\arabic*.}}

\textit{Legends: MC - Multiple Choice, SC - Single Choice, OE - Open-ended}

\textbf{Participant Consent:} \textit{Participant's consent to participate in the study.}

\textbf{Participant Background:} \textit{Age, Gender, Occupation, Education, Annual Income, Area of Living (Urban, Semi-Urban, Rural)}

\textbf{Filtering PHI Users and Non-Users:} \textit{Do you own or have you ever owned a PHI wearable? (Yes/No)}

\textbf{PHI Awareness and Usage (Users)}
\begin{enumerate}[resume]
  \item Types of PHI wearables owned or previously owned (OE).
  \item Reasons for using PHI wearables (MC + OE).
  \item Health markers you track with the PHI wearable (MC + OE).
  \item Frequency of PHI wearable use (SC).
  \item Which health markers are available to you through the PHI wearable (MC + OE)?
  \item Have you ever changed the PHI wearable you use? (Yes/No)
\end{enumerate}

\textbf{PHI Understanding, Utility, and Trust (Users)}
\begin{enumerate}[resume]
  \item What do you understand about the data and analysis produced by your PHI wearable? (OE)
  \item How much of your PHI health history is viewable/usable to you? (OE)
  \item Does your PHI health history ever help you during doctor visits? (Yes/No) 
  \item Please briefly explain your choice in the previous question. (OE)
  \item Do you trust the accuracy of data collected through the PHI wearable? (Yes/No)
  \item Please briefly explain your choice regarding trust in the previous question. (OE)
  \item Has there ever been a mismatch between how you feel and physiological values recorded by your PHI wearable? (Yes/No/Others)
  \item Please explain instances of such mismatches, if any. (OE)
  \item If you have had problems, did you continue using the PHI wearable afterwards? (Yes/No)
  \item If yes, please briefly explain why you continued using it. (OE)
  \item Are you interested in answering questions about your PHI usage experience further in a short online interview? (Yes/No)
\end{enumerate}

\textbf{Understanding Experience with PHI (Non-Users)}
\begin{enumerate}[resume]
  \item Has anyone you know (friends/family/acquaintances) used or ever used PHI wearables? (Yes/No; please expand if relevant (OE))
  \item Has anyone you know (friends/family/acquaintances) ever recommended that you use a PHI wearable? (Yes/No)
  \item Are you interested in answering questions about your experience further in a short online interview? (Yes/No)
\end{enumerate}

\textbf{Collecting Contact Information for Interested Participants}
\begin{enumerate}[resume]
  \item If you are willing to be contacted for follow-up, please provide your email address:\\[4pt]
    \textit{Email:} \rule{8cm}{0.4pt}
\end{enumerate}

\end{tcolorbox}

\section{Interview Guide (RQ1)}
\label{appendix:interview_rq1}

\begin{tcolorbox}[enhanced,breakable,
                  colback=white,colframe=black,boxrule=0.6pt,arc=1pt,
                  left=6pt,right=6pt,top=6pt,bottom=6pt,
                  before skip=6pt,after skip=6pt]
\footnotesize
\setlist[itemize]{noitemsep, topsep=0pt, left=12pt}
\setlist[enumerate]{noitemsep, topsep=2pt, left=0pt, label=\textbf{Q\arabic*.}}

\textbf{Participant Consent:} \textit{Participant's consent to participate in the study.}

\textbf{Everyday Care Practices}
\begin{enumerate}
    \item Describe your typical day. At what points, if any, do you find yourself thinking about your health or your body’s signals?
    \item What is your immediate response when you notice physical changes or discomfort?
    \item How do you currently organize health-related information, such as appointments or test results?
    \item Who do you first discuss health concerns with, and what is that interaction typically like?
    \item Who leads health-related decisions or habit changes within your household?
    \item How do you respond to health advice or suggestions from family and friends?
    \item How do you distinguish between private health information and data you are comfortable sharing?
    \item What information is necessary for family support, and what feels like an overshare?
    \item Describe the process of explaining health metrics or clinical notes to family members.
    \item How do you decide whether a health signal is a significant problem or a minor occurrence?
    \item How do embodied cues (body-feeling) influence your help-seeking compared to digital data?
\end{enumerate}

\textbf{PHI Awareness and Usage (Users)}
\begin{enumerate}[resume]
  \item What types of wearables have you owned?
  \item What is the purpose of using wearables?
  \item How often do you track your metrics?
  \item What happens if you miss tracking? Does it affect you in any way?
  \item How easy/difficult is it to view/export/utilize Health Information/Analysis?
\end{enumerate}

\textbf{PHI Understanding, Utility, and Trust (Users)}
\begin{enumerate}[resume]
  \item What is your trust in the data from the wearable?
  \item How well do you understand the tracked data?
  \item How do (if) you use this information for proactive care?
  \item How do you deal with inaccuracies or mismatches of Health data?
  \item Have you had any other barriers to the effective adoption and utilization of wearables for proactive care?
  \item What is the reason for the continued use of the wearable?
\end{enumerate}

\textbf{Understanding Experience with PHI (Non-Users)}
\begin{enumerate}[resume]
  \item Why are you uninterested in using PHI?
  \item What are the barriers towards effective adoption of PHI or proactive care?
  \item What is your perception of overcoming this barrier?
  \item What could make PHI better aligned with your lived realities?
\end{enumerate}

\end{tcolorbox}

\section{Stakeholder Interview Guide (RQ2)}
\label{appendix:interview_rq2}

\begin{tcolorbox}[enhanced,breakable,
                  colback=white,colframe=black,boxrule=0.6pt,arc=1pt,
                  left=6pt,right=6pt,top=6pt,bottom=6pt,
                  before skip=6pt,after skip=6pt]
\footnotesize
\setlist[itemize]{noitemsep, topsep=0pt, left=12pt}
\setlist[enumerate]{noitemsep, topsep=2pt, left=0pt, label=\textbf{Q\arabic*.}}

\textbf{Participant Consent:} \textit{Participant's consent to participate in the study.}

\textbf{Older Adults}
\begin{enumerate}[resume]
  \item What types of PHI are you comfortable sharing?
  \item How frequently would you like to share?
  \item Which stakeholders do you want to share it with?
  \item What is your perception of privacy when it comes to PHI?
  \item Which part of PHI are you uncomfortable sharing, and with whom?
  \item Do you wish to see PHI of your adult children? If yes, how frequently?
  \item In what way do you want your PHI or your loved ones' PHI represented?
\end{enumerate}

\textbf{Adult Children (informal caregivers of Older Adults)}
\begin{enumerate}[resume]
  \item What level of information is required to support the older adults better?
  \item How frequently do you need this information?
  \item What is your perception of privacy on PHI?
  \item Which of your PHI are you open to sharing with your parents?
  \item What other support do you require apart from PHI of your parents?
\end{enumerate}

\textbf{Doctors}
\begin{enumerate}[resume]
  \item Which PHI information is useful when patients visit?
  \item What do you think about patients' PHI being part of their health history?
  \item Is more information better/worse for diagnosis or treatment?
  \item What is the acceptable amount of PHI history before it is extra work for the doctors?
  \item Do you have any experience in using PHI for diagnosis or treatment?
  \item If yes, how did it help, how was the process, what can be done to make it better (if applicable)?
\end{enumerate}

\end{tcolorbox}

\section{Survey Questionnaire (RQ2)}
\label{appendix:survey_rq2}

\begin{tcolorbox}[enhanced,breakable,
                  colback=white,colframe=black,boxrule=0.6pt,arc=1pt,
                  left=6pt,right=6pt,top=6pt,bottom=6pt,
                  before skip=6pt,after skip=6pt]
\footnotesize
\setlist[itemize]{noitemsep, topsep=0pt, left=12pt}
\setlist[enumerate]{noitemsep, topsep=2pt, left=0pt, label=\textbf{Q\arabic*.}}

\textit{Legends: MC - Multiple Choice, SC - Single Choice, OE - Open-ended}

\textbf{Participant Consent:} \textit{Participant's consent to participate in the study.}

\textbf{Participant Background:} \textit{Age, Gender, Education, Technology Proficiency}

\textbf{Filtering Android and iOS Users:} \textit{What type of smartphone do you use? (Android/iOS)}

\textbf{Apple Health Questions}
\begin{enumerate}[resume]
  \item How frequently do you use Apple Health? (SC)
  \item How often have you used Apple Health sharing? (SC)
\end{enumerate}

\textbf{Apple Health Sharing User's Awareness, Understanding and Use}
\begin{enumerate}[resume]
  \item With whom can you use the health sharing feature? (SC)
  \item What is the understanding of the sharing feature you use in Apple Health? (SC)
  \item Which of the following actions can you perform in your health sharing feature? (MC)
  \item In your own words, please share your experience of using the sharing feature. (OE)
\end{enumerate}

\textbf{Apple Health Sharing Non-Users' Understanding and Motivation}
\begin{enumerate}[resume]
  \item Are you interested in sharing your health information? If yes, what are you interested in sharing? (SC + OE)
  \item How would you like to share that information? (SC + OE)
  \item What use do you envision in sharing health data? (N/A if not interested in Sharing) (OE)
  \item If you heard about the sharing feature in Apple Health, why don't you use it? (N/A if not applicable) (OE)
\end{enumerate}

\textbf{Android Health Questions}
\begin{enumerate}[resume]
  \item How frequently do you use any health application such as Google Fit, Fitbit, Samsung Health, etc.? (SC)
  \item How often have you used the sharing feature of any such health application? (SC)
\end{enumerate}

\textbf{Android Health Sharing User's Awareness, Understanding and Use}
\begin{enumerate}[resume]
  \item With whom can you use the health sharing feature? (SC)
  \item What is the understanding of the sharing feature you use in your health application? (SC)
  \item Which of the following actions can you perform in your health sharing feature? (MC)
  \item In your own words, please share your experience of using the sharing feature. (OE)
\end{enumerate}

\textbf{Android Health Sharing Non-Users' Understanding and Motivation}
\begin{enumerate}[resume]
  \item Are you interested in sharing your health information? If yes, what are you interested in sharing? (SC + OE)
  \item How would you like to share that information? (SC + OE)
  \item What use do you envision in sharing health data? (Please write N/A if not interested in Sharing) (OE)
  \item If you heard about the sharing features in the health application you use, why don't you use them? (Please type N/A if not applicable) (OE)
\end{enumerate}

\end{tcolorbox}

\end{document}